\documentclass[11pt]{article}
\usepackage{amssymb}
\usepackage{amsmath}
\usepackage{amstext}
\usepackage{graphicx,epsfig}
\usepackage{epsfig}
\usepackage{verbatim} 
\usepackage{fancybox}
\usepackage{color}
\usepackage{ulem}
\usepackage{enumitem}
\usepackage{subfigure}
\usepackage{bbm}
\usepackage{parskip}
\usepackage{dsfont}
\usepackage[numbers,sort&compress]{natbib}
\usepackage{bm}
\usepackage[dvipsnames]{xcolor}

\usepackage{hyperref} %Automatically links \label and \ref commands; Always load last
\usepackage[all]{hypcap} %Link navagates to top of figure instead of caption (below fig)
\hypersetup{
    colorlinks=true,       % false: boxed links; true: colored links
    linkcolor=red,          % color of internal links
    citecolor=blue,        % color of links to bibliography
    filecolor=magenta,      % color of file links
    urlcolor=blue           % color of external links
}

\usepackage{myterms}

\newcommand{\PT}{\text{\sc pt}}

\numberwithin{equation}{section}

\begin{document}
%
%\maketitle
~
\vspace{1truecm}
\renewcommand{\thefootnote}{\fnsymbol{footnote}}
\begin{center}
{\huge \bf{
Probing the Electroweak Phase Transition 
\\ \vspace{0.2cm} 
with Higgs Factories and Gravitational Waves 
}}
\end{center}

\vspace{1truecm}
\thispagestyle{empty}
\centerline{\Large Peisi Huang,${}^{\rm a,b}$\footnote{\tt peisi@uchicago.edu} Andrew J. Long,${}^{\rm c}$\footnote{\tt andrewjlong@kicp.uchicago.edu} and Lian-Tao Wang${}^{\rm a,c}$\footnote{\tt liantaow@uchicago.edu}}
\vspace{.7cm}

\centerline{\it ${}^{\rm a}$Enrico Fermi Institute, University of Chicago, Chicago, Illinois 60637, USA.}

\vspace{.2cm}

\centerline{\it ${}^{\rm b}$High Energy Physics Division,  Argonne National Laboratory, Argonne, Illinois 60439, USA}
%\centerline{\it }
\vspace{.2cm}
\centerline{\it$^{\rm c}$Kavli Institute for Cosmological Physics, University of Chicago, Chicago, Illinois 60637, USA}

\vspace{.5cm}
\begin{abstract}
\vspace{.03cm}
\noindent
After the discovery of the Higgs boson, understanding the nature of electroweak symmetry breaking and the associated electroweak phase transition has become the most pressing question in particle physics.  
Answering this question is a priority for experimental studies.  
Data from the LHC and future lepton collider-based Higgs factories may uncover new physics coupled to the Higgs boson, which can induce the electroweak phase transition to become first order.  
Such a phase transition generates a stochastic background of gravitational waves, which could potentially be detected by a space-based gravitational wave interferometer.  
In this paper, we survey a few classes of models in which the electroweak phase transition is strongly first order.  
We identify the observables that would provide evidence of these models at the LHC and next-generation lepton colliders, and we assess whether the corresponding gravitational wave signal could be detected by eLISA.  
We find that most of the models with first order electroweak phase transition can be covered by the precise measurements of Higgs couplings at the proposed Higgs factories.  
We also map out the model space that can be probed with gravitational wave detection by eLISA.  
\end{abstract}

\newpage

\begingroup
\setlength{\parskip}{0cm}
\hypersetup{linkcolor=black}
\tableofcontents
\endgroup

%\newpage
\renewcommand*{\thefootnote}{\arabic{footnote}}
\setcounter{footnote}{0}

%==================================
% INTRODUCTION
%==================================
\section{Introduction}\label{sec:Introduction}

%==========
The discovery of the Higgs boson completes the list of particles of the Standard Model.  
However, there are many open questions regarding the dynamics of electroweak symmetry breaking.  
Addressing these questions has been a driving force in both theoretical and experimental explorations of the energy frontier. 

%==========
One of the most outstanding problem is the nature of the electroweak phase transition.  
Currently, we have measured with precision the size of the Higgs vacuum expectation value (vev) and the mass of the Higgs boson.  
However, we know very little about the shape of the Higgs potential beyond that.  
The Standard Model defines its Higgs potential with only renormalizable terms -- the so-called ``Mexican" hat form.  
In this case the electroweak symmetry is smoothly restored via a continuous cross-over as the temperature is raised above the electroweak scale \cite{DOnofrio:2015mpa}.  
However, new physics can modify the nature of the phase transition, possibly turning it into an abrupt first-order phase transition.  
On the one hand, such new physics can be searched for directly at current and future colliders.  
On the other hand, a first order electroweak phase transition in the early universe would generate a stochastic background of gravitational waves that can be searched for with interferometers \cite{Caprini:2015zlo}.  
Discovering such a gravitational wave signal would establish the electroweak phase transition as a new milestone in our understanding of the early universe.  
It will advance our knowledge into an epoch significantly earlier than nucleosynthesis.  

%==========
A first order electroweak phase transition requires a significant deviation away from the renormalizable Higgs potential, which implies the presence of new physics close to the weak scale.  
We can look for such new particles directly, such as at the LHC.  
Even though it can be powerful in certain cases, the reach of this approach is limited.  
The new physics can be weakly coupled, and the searches at the LHC suffers from large background.  
At the same time, the most model independent effect of such new physics is the induced deviation in the Higgs couplings \cite{Noble:2007kk, Katz:2014bha}.  
Measuring such couplings precisely, and uncovering potential new physics, is a major physics goal of proposed Higgs factories.  
In this paper, we will focus on the potential of probing new physics associated with electroweak symmetry breaking at these facilities. 

%==========
If a first order electroweak phase transition occurred in the early universe, the collision of bubbles and damping of plasma inhomogeneities would have generated a stochastic background of gravitational waves.  
The frequency of these waves is relatively model independent, being related to the scale of the cosmological horizon at the time of the phase transition.  
Therefore today we expect the waves to have redshifted into the milli-Hertz range.  
This potential signal is impossible to probe with ground-based gravitational wave interferometers like AdvLIGO due to seismic noise.  
However, the signal is ideal for a space-based interferometer like eLISA \cite{Caprini:2015zlo} with arm lengths of order millions of kilometers.  
In this paper, we assess the possibility of using gravitational waves as probes of a first order electroweak phase transition and the complementarity of this technique with the collider searches.  

%==========
There are a number of ways in which new particles may cause the electroweak phase transition to become first order.  
In general a first order phase transition can occur if the Higgs effective potential is modified from its Standard Model form so as to develop a potential energy barrier separating the phases of broken and unbroken electroweak symmetry.  
As discussed in \rref{Chung:2012vg}, there are three general model classes in which the barrier can arise.  
First, if the new degrees of freedom are scalar fields that participate in the electroweak phase transition (their vev changes at the same time as the Higgs) then {\it tree-level interactions} with the Higgs field can make some regions of field space energetically dis-favorable and lead to a barrier.  
Second, the presence of new particles coupled to the Higgs boson affects the running of the Higgs mass parameter and self-coupling.  
Then the barrier can arise by virtue of {\it quantum effects}.  
Third, if the new particles are present in the early universe plasma and acquire their mass (at least partially) from the Higgs field, then a barrier can arise via {\it thermal effects}.  
This can be understood as a tradeoff between minimizing the energy, represented by the tree-level Higgs potential, and maximizing the entropy, which prefers the Higgs field to take on values where particles in the plasma are light.  
Then, this third case can be further divided into two categories:  a barrier arising from light scalars via the thermal cubic term $\sim(m^2)^{3/2} T$ and a barrier arising from heavy particles that get their mass predominantly from a large coupling with the Higgs field.   

%==========
In this paper, we survey a number of simplified models that demonstrate the basic ingredients necessary for a first order electroweak phase transition.  
We focus on four models in which the SM is extended, respectively, to include a real scalar singlet, a scalar doublet, heavy chiral fermions, and varying Yukawa couplings.  
This set of models exemplifies all of the different phase transition model classes, enumerated above.  

%==================================
% MODELS
%==================================
\section{Models}\label{sec:Models}

%==========
In each of the models discussed here, the Higgs field is represented by $\Phi(x)$, and the Standard Model Lagrangian contains
\begin{align}
	\Lcal_{\SM} \supset \bigl( D_{\mu} \Phi \bigr)^{\dagger} \bigl( D^{\mu} \Phi \bigr) - m_0^2 \Phi^{\dagger} \Phi - \lambda_h \bigl( \Phi^{\dagger} \Phi \bigr)^2
	\per
\end{align}
In calculating the scalar effective potential we write $\langle \Phi(x) \rangle = (0 \, , \, \phi_{h}/\sqrt{2})$ with $\phi_h$ real.  
The vacuum spontaneously breaks the electroweak symmetry, $\phi_h = v$ with $v \simeq 246 \GeV$.  
The Higgs mass is denoted as $M_h$, and it takes the value $M_h \simeq 125 \GeV$.  

%----------------------------------------------------------------
% Real Scalar Singlet
%----------------------------------------------------------------
\subsection{Real Scalar Singlet}\label{sec:Singlet}

%==========
First, we add to the SM a real scalar field $S(x)$, which is a singlet under the SM gauge group.  
This is probably the simplest extension of the Higgs sector of the SM.  
At the same time, due the lack of other interactions, it gives rise to the most independent signal. 

%==========
The most general renormalizable Lagrangian is written as 
\begin{align}\label{eq:singlet_L}
	\Lcal & = \Lcal_{\SM} + \frac{1}{2} \bigl( \partial_{\mu} S \bigr) \bigl( \partial^{\mu} S \bigr) - t_s S - \frac{m_s^2}{2} S^2 - \frac{a_s}{3} S^3 - \frac{\lambda_s}{4} S^4 - \lambda_{hs} \Phi^{\dagger} \Phi S^2 - 2 a_{hs} \Phi^{\dagger} \Phi S
	\per
\end{align}
Without loss of generality, we can set $t_s=0$.  
Since the new scalar is a singlet, it only interacts with the Standard Model via the Higgs portal, $\Phi^{\dagger} \Phi S^2$ and $\Phi^{\dagger} \Phi S$.  
The electroweak phase transition and collider phenomenology in this model, sometimes called the {\rm xSM}, have been studied extensively; see {\it e.g.} \rref{Kotwal:2016tex} and references therein.  
The gravitational wave signal in related models has been studied recently by Refs.~\cite{Leitao:2015fmj, Huber:2015znp, Huang:2015izx, Huang:2016odd, Chala:2016ykx, Kakizaki:2015wua, Hashino:2016rvx}

%==========
There is no single reason why this model admits a first order electroweak phase transition.  
In fact different limits of this simple model exhibit each of the phase transition model classes that were identified in \rref{Chung:2012vg}.  
Most notably, the tree-level interactions play a significant role in most of the parameter space.  
During the electroweak phase transition, the singlet $v_{s}$ need not remain fixed.  
If $v_{s}$ changes along with $v$, then the Higgs portal terms, $\lambda_{hs} \Phi^{\dagger} \Phi S^2 $ and $a_{hs} \Phi^\dagger \Phi S$, can give rise to a barrier in the effective potential, and the phase transition is first order.  

%==========
After electroweak symmetry breaking $\langle \Phi \rangle = (0 \, , \, v/\sqrt{2})$, and generically we expect the singlet field to acquire a vacuum expectation value as well, $\langle S \rangle = v_s$.  
Then, the Higgs portal operators allow the Higgs and singlet fields to mix. 
The mixing angle $-\pi/4 \leq \theta \leq \pi/4$ satisfies
\begin{align}\label{eq:singlet_mixing}
	\sin 2 \theta = \frac{4v(a_{hs} + \lambda_{hs} v_s)}{M_h^2-M_s^2}
\end{align}
where $M_h \simeq 125 \GeV$ is the physical Higgs boson mass, and $M_s$ is the physical mass of the singlet.  

%==========
Interactions between the Higgs boson and the singlet scalar affect the coupling of the Higgs to the Z-boson.  
Writing the effective $hZZ$ coupling as $g_{hZZ}$, we define the fractional deviation from the SM value as\footnote{In an earlier version of this article, we used the notation $\delta Z_h = 1 - g_{hZZ} / g_{hZZ,\SM}$ to represent the fractional change in the $hZZ$ coupling.  To avoid confusion with \rref{Craig:2013xia}, where $\delta Z_h$ denotes the Higgs field wavefunction renormalization, we have changed the notation in \eref{eq:singlet_dZh}.  We have the relations $\delta g_{hZZ} = - \delta Z_h/2$, and the fractional change in the $hZ$ production cross section is given by $\sigma_{hZ} / \sigma_{hZ}^{\SM} - 1 = 2 \, \delta g_{hZZ} = - \delta Z_h$.  }  
\begin{align}\label{eq:dZh_def}
%	\delta Z_h \equiv \left. 1 - \frac{g_{hZZ}}{g_{hZZ, \SM}} \right|_{s=(250 \GeV)^2}
	\delta g_{hZZ} \equiv \left. \frac{g_{hZZ}}{g_{hZZ, \SM}} - 1 \right|_{s=(250 \GeV)^2}
\end{align}
where the couplings are evaluated at a center of mass energy $s = (250\GeV)^2$.  
We calculate $\delta g_{hZZ}$ as
\begin{align}\label{eq:singlet_dZh}
	\delta g_{hZZ}
	\approx & 
	\bigl( \cos \theta - 1 \bigr) 
%	- \frac{1}{2} \frac{|a_{hs} + \lambda_{hs} v_{s}|^2}{16 \pi^2} I_B(M_h^2; M_h^2, M_s^2) 
	-2 \frac{|a_{hs} + \lambda_{hs} v_{s}|^2}{16 \pi^2} I_B(M_h^2; M_h^2, M_s^2) 
	\\ &
%	- \frac{1}{2} \frac{|\lambda_{hs}|^2 v^2}{16 \pi^2} I_B(M_h^2; M_s^2, M_s^2) 
	- \frac{|\lambda_{hs}|^2 v^2}{16 \pi^2} I_B(M_h^2; M_s^2, M_s^2) 
	+ 0.006 \left( \frac{\lambda_{3}}{\lambda_{3,\SM}}-1 \right) 
	\nonumber
	\per
\end{align}
The first term arises from the tree-level Higgs-singlet mixing (\ref{eq:singlet_mixing}).  
This is typically the dominant contribution to $\delta g_{hZZ}$.  
At the one-loop order, the singlet contributes to the wavefunction renormalization of the Higgs.  
This gives rise to the second and third terms in \eref{eq:singlet_dZh}.  
We have generalized the calculation in Refs.~\cite{Craig:2013xia,Curtin:2014jma}, to allow for the cases without a $\Zbb_2$ symmetry.  
The bosonic loop function is given by~\cite{Fan:2014axa}
\begin{equation}\label{eq:IB_def}
	I_B(p^2; m_1^2,m_2^2) = \int_0^1 \! \ud x \ \frac{x (1-x)}{x (1-x) p^2 - x m_1^2 - (1-x) m_2^2}
	\per 
\end{equation} 
The wavefunction renormalization terms are typically sub-dominant, except for the $\Zbb_2$ limit (discussed in \sref{sec:Z2}) where $\theta = a_{hs} = v_{s} = 0$ and the $|\lambda_{hs}|^2 v^2$ term is dominant.    

%==========
The fourth term in \eref{eq:dZh_def} also arises at the one-loop order.  
As recognized in \rref{McCullough:2013rea} this term appears when the Higgs trilinear coupling $\lambda_{3}$ deviates from its SM value $\lambda_{3,\SM}$.  
The effect on $\delta g_{hZZ}$ depends on the center of mass energy, and for $\sqrt{s} = 250 \GeV$ the prefactor evaluates to $0.006$~\cite{McCullough:2013rea}.  
The cubic self-coupling of the mass eigenstate Higgs ($hhh$) is calculated as 
\begin{align}\label{eq:singlet_lam3}
	\lambda_{3} 
	& 
	= \bigl( 6 \lambda_{h} v \bigr) \cos^3 \theta 
	+ \bigl( 6 a_{hs} + 6 \lambda_{hs} v_s \bigr) \sin \theta \cos^2 \theta
	+ \bigl( 6 \lambda_{hs} v \bigr) \sin^2 \theta \cos \theta 
	+ \bigl( 2 a_{s} + 6 \lambda_{s} v_{s} \bigr) \sin^3 \theta 
	\per
\end{align}
In the Standard Model we have $\lambda_{3} = \lambda_{3,\SM} \equiv 3 M_h^2/v \simeq 191 \GeV$.  
The last term in $\lambda_{3}$ arises from a three-vertex, one-loop graph.  
As we will see, models exhibiting a strongly first order phase transition, typically have an $O(1)$ deviation in $\lambda_3$, and therefore, this effect on $\delta g_{hZZ}$ can be sizable. 

%==========
If the singlet is sufficiently light, $M_s < M_h  / 2 \simeq 62.5 \GeV$, the Higgs decay channel $h \to SS$ opens.  
This decay contributes to the Higgs invisible width.  
The invisible width is calculated as~\cite{Chen:2014ask}
\begin{align}
	\Gamma_{\rm inv} = \Gamma(h \to SS) = \frac{\lambda_{211}^2}{32\pi M_h} \sqrt{ 1 - \frac{4M_s^2}{M_h^2} }
\end{align}
where 
\begin{align}
	\lambda_{211} 
	& 
	= \bigl( 2 a_{hs} + 2 \lambda_{hs} v_{s} \bigr) \cos^3 \theta 
	+ \bigl( 4 \lambda_{hs} v - 6 \lambda_{h} v \bigr) \sin \theta \cos^2 \theta \nn
	& \quad 
	+ \bigl( 6 \lambda_{s} v_{s} + 2 a_{s} - 4 \lambda_{hs} v_{s} - 4 a_{hs} \bigr) \sin^2 \theta \cos \theta 
	+ \bigl( - 2 \lambda_{hs} v \bigr) \sin^3 \theta
	\per
\end{align}
is the effective tri-linear coupling of the mass eigenstates.  
If the invisible channel is open, the branching fraction is typically so large as to be excluded already by LHC limits, ${\rm BR}_{\rm inv} \lesssim 30\%$ \cite{Aad:2015pla, CMS:2016jjx}.  Therefore we require $M_s > M_h/2$.  

%==========
In general the model has $7$ parameters corresponding to the potential terms in \eref{eq:singlet_L} with $t_s=0$.  
Two parameters can be exchanged for the Higgs mass and vev, leaving $5$ free parameters.  
In \sref{sec:Results} we present the main result for this model, which entails a scan over the $5$-dimensional parameter space.  
In the following sub-sections we discuss a couple of special limiting cases of this model.  
Even though they do not represent generic models with first order electroweak phase transition, they give rise to different predictions.  
We include them in our discussion for completeness. 

%----------------------------------------------------------------
% Z2-Symmetric Limit
%----------------------------------------------------------------
\subsubsection{$\Zbb_2$-Symmetric Limit}\label{sec:Z2}

%==========
We impose the $\Zbb_2$ discrete symmetry under which the singlet is odd, $\phi_s \to -\phi_s$, and the Standard Model fields are even.  
In terms of the singlet Lagrangian (\ref{eq:singlet_L}) the symmetry enforces
\begin{align}\label{eq:singlet_Z2}
	t_s = 0 \, , \quad
	a_s = 0 \, , \quad \text{and} \quad
	a_{hs} = 0
	\per
\end{align}
We also require that the $\Zbb_2$ symmetry is not broken spontaneously, and thus $v_s=0$.  
The only interaction between the Standard Model and the singlet is through the Higgs portal, $\lambda_{hs} \Phi^{\dagger} \Phi S^2$.  

%==========
The $\Zbb_2$ symmetry forbids a mixing between the Higgs and singlet fields.  
In the absence of mixing, modifications to the $hZZ$ coupling (\ref{eq:singlet_dZh}) first arise at the one-loop level.  
The tree level modifications to the trilinear coupling (\ref{eq:singlet_lam3}) are also suppressed in this limit, and therefore, only the $|\lambda_{hs}|^2$ term contributes to $\delta g_{hZZ}$.  
Thus we expect that this corner of parameter space can evade constraints on $\delta g_{hZZ}$ from future colliders as discussed later.  
As such, this model has been identified as a ``worst case scenario'' for finding evidence of a first order EW phase transition at colliders \cite{Ashoorioon:2009nf, Curtin:2014jma}.  

%==========
Despite the vanishing mixing, the electroweak phase transition may still be first order.  
Morally speaking, the Higgs-singlet mixing is a ``local'' property of the theory, related to the behavior of small fluctuations about the vacuum, but the nature of the phase transition depends also up ``global'' properties of the theory, {\it e.g.} whether the theory admits other metastable vacua.  
The presence or absence of such metastable vacua is not {\it directly} related to the mixing at the true vacuum.  
Two specific scenarios have been studied.  
If the Higgs portal coupling is sufficiently large, the singlet can affect the running of the Higgs self-coupling, which may induce a barrier in the effective potential \cite{Espinosa:2007qk}.  
Alternatively, the Higgs portal interaction, $\lambda_{hs} \phi_h^2 \phi_s^2$ with $\lambda_{hs} > 0$, may give rise to a barrier in the effective potential at tree-level when the phase transition passes from a vacuum with $\phi_h=0$ and $\phi_s = v_s(T)$ to a vacuum with $\phi_h = v(T)$ and $\phi_s = 0$.  

%----------------------------------------------------------------
% Unmixed Limit
%----------------------------------------------------------------
\subsubsection{Unmixed Limit}\label{sec:Unmixed}

%===============
The tree-level Higgs-singlet mixing (\ref{eq:singlet_mixing}) vanishes when we take
\begin{align}\label{eq:no_mixing}
	a_{hs} + \lambda_{hs} v_{s} = 0 
	\per
\end{align}
Unlike the $\Zbb_2$ symmetric limit of \eref{eq:singlet_Z2}, the choice of parameters in \eref{eq:no_mixing} is not associated with any enhanced symmetry.  
Significant fine-tuning among tree-level parameters is necessary to reach this limit.  
Furthermore, such a tuning is not technically natural; the mixing is induced radiatively.  
At the one-loop order, the induced mixing is $\propto \lambda_{hs} v (a_{s} + \lambda_{hs} v_{s})$.  
Otherwise, the first order electroweak phase transition and collider phenomenology is similar to the $\Zbb_2$ case.  

%===============
In the unmixed limit, the singlet can be pair produced through an off-shell Higgs via the $hSS$ coupling $\lambda_{hs} v$, and can decay to $hh$, $WW$, and $ZZ$ final states via the radiatively generated mixing.  
Then by the Goldstone equivalence theorem, the singlet decay branching ratios for the $hh$, $WW$, and $ZZ$ channels are 25$\%$, 50$\%$, and 25$\%$ respectively.  
Those final states contribute to a multi-lepton, multi-jet signature, which can be probed at the LHC.  
With a large $WW$ branching ratio, the 4$W$ channel leads to a same-sign dilepton with multiple jets (zero b-jet) final state.  
The background processes for this channel include $t\bar{t}$, $t\bar{t}W$, $t\bar{t}Z$, $WZ$, and same sign $WW$ plus jets.  
All backgrounds except same sign $WW$ plus jets, can be estimated from the $t\bar{t}h$ searches in the same sign dilepton with at least two b-jets channel, by replacing the b-tagging with a b-jet veto~\cite{Aad:2015iha}.  
We assume the b-tagging efficiency is 70$\%$.  
For the same sign $WW$ plus jets, we include both single parton scattering and double parton scattering~\cite{Melia:2010bm}, and assume a 90$\%$ acceptance to account for the lepton efficiency, and kinematics.  
We assume the same 90$\%$ acceptance for signal as well.  
Then at the HL-LHC, we expect a 2$\sigma$ significance for $\sigma (p p \rightarrow SS) \sim 1.8 \fb$, which corresponds to $\lambda_{hs} \sim 1$, and $M_S \sim 200 \GeV$~\cite{CEPC-SPPCStudyGroup:2015csa}.

%----------------------------------------------------------------
% Scalar Doublet (Stop-Like)
%----------------------------------------------------------------
\subsection{Scalar Doublet (Stop-Like)}\label{sec:Stop_Like}

%==========
In this section, we go beyond the singlet to consider new particles in non-trivial representations of the SM gauge group.  
Some of the simplest cases are obtained by introducing $\SU{2}_L$ scalar doublets and singlets with $\U{1}_Y$ charge. 
Perhaps the most well-known example is the MSSM stop. 
However, the light stop scenario is very restricted and, at least in simple cases, it can not give rise to a first order electroweak phase transition without running afoul of collider constraints \cite{Curtin:2012aa, Cohen:2012zza, Carena:2012np}, see also~\cite{Huang:2012wn}.  
Many of these constraints are a consequence of the supersymmetry.  
For example, the scalar top partner must to be colored and hence the stop is subject to stringent limits from collider searches. To avoid the collider constraints, models like folded SUSY have been proposed~\cite{Burdman:2006tz}, in which the stops can still solve the hierarchy problem, but are not colored.  
In the following, we consider a similar stop-like model.  
The new particles are taken to have the same electroweak gauge quantum numbers as the stop, but they are not colored.  
In addition, their couplings are not subject to the constraints of supersymmetry.  

%==========
We extend the SM to include $n_f = 3$ copies (flavors) of scalar doublets and complex scalar singlets.  
We will denote the doublets and singlets as $\tilde{Q}_{i} = (\tilde{u}_{i} \, , \, \tilde{d}_{i})^T$ and $\tilde{U}_{i}$ where the index $i$ runs from $1$ to $n_f$.  
In order to mimic the interactions of colored squarks, we require the Lagrangian to respect the global $\SU{n_f}$ symmetry, under which the $\tilde{Q}_{i}$ and $\tilde{U}_{i}$ transform in the fundamental representation, and the SM fields are invariant.  
Notice that we have used a SUSY-like notation to indicate the electroweak gauge quantum numbers, but no SUSY relations are implied.  

%==========
With the new stop-like particle content, the scalar potential can be written as 
\begin{align}\label{eq:VQU}
	V & = \frac{1}{2} m_0^2 \phi_h^2 + \frac{\lambda_h}{4} \phi_h^4 \\
	& \quad + m_{Q}^2 \bigl( |\tilde{u}|^2 + |\tilde{d}|^2 \bigr) + m_{U}^2 |\tilde{U}|^2 + \lambda_{Q} \bigl( |\tilde{u}|^2 + |\tilde{d}|^2 \bigr)^2 + \lambda_{U} \bigl( |\tilde{U}|^2 \bigr)^2 \nn
	& \quad + \lambda_{QU} \bigl( |\tilde{u}|^2 + |\tilde{d}|^2 \bigr) |\tilde{U}|^2 + \frac{\lambda_{hU}}{2} \phi_h^2 |\tilde{U}|^2  \nn
	& \quad + \frac{\lambda_{hQ}}{2} \bigl( |\tilde{u}|^2 + |\tilde{d}|^2 \bigr) \phi_h^2 + \frac{\lambda_{hQ}^{\prime}}{2} |\tilde{u}|^2 \phi_h^2 + \frac{\lambda_{hQ}^{\prime \prime}}{2} |\tilde{d}|^2 \phi_h^2 \nn
	& \quad + \bigl[ \frac{a_{hQU}}{\sqrt{2}} \tilde{u} \phi_h \tilde{U}^{\ast} + \hc \bigr]
	\nonumber
	\per
\end{align}
The sum over $i = 1, \cdots, n_f$ flavors has been suppressed.  
In general the model has $12$ parameters, but $2$ of these can be exchanged for the Higgs mass and vev, leaving $10$ free parameters.  
Additionally, we will later assume a universal dimensionless coupling, $\lambda_{Q} = \lambda_{U} = \lambda_{UQ} = \cdots \equiv \lambda$, which reduces the free parameters to four:  $\{ m_Q^2, m_U^2, \lambda, a_{hQU} \}$.  
We present the results of a parameter-space scan in \sref{sec:Results}.  

%==========
In the well-known light stop scenario of the MSSM \cite{Carena:1996wj}, the electroweak phase transition can become first order due to the presence of these scalar particles in the plasma.  
Their contribution to the Higgs thermal effective potential (background-dependent free energy density) goes as $V_{\rm eff} \sim - N_{c} [m_{\tilde{t}}(\phi_h,T)^2 ]^{3/2} T$ where $N_{c} = 3$ is the number of colors, and the effective stop mass $m_{\tilde{t}}(\phi_h,T)$ depends on the background Higgs field $\phi_h$ and the plasma temperature $T$ (daisy correction).  
This non-analytic term in $V_{\rm eff}$ arises only for relativistic bosonic fields, due to the non-analyticity of the Bose-Einstein distribution function at $E/T = 0$.  
In a regime where the stop mass can be approximated as $m_{\tilde{t}}(\phi_h)^2 \approx y_t^2 \phi_h^2$, the effective potential acquires a cubic term, $V_{\rm eff} \sim N_{c} y_{t}^{3} \phi_h^3 T$, which can provide the requisite barrier for a first order phase transition.  
In our stop-like model, the same thermal effects can give rise to a first order electroweak phase transition.  
However, since we do not impose the SUSY relations, $y_t$ is replaced by some combination of the quartic couplings $\lambda$ in \eref{eq:VQU}.  
In principle, $\lambda$ can be larger than $y_t$, which increases the height of the potential energy barrier and strengthens the first order the phase transition.  
For simplicity, we neglect the daisy resummation, {\it i.e.} the one-loop temperature-dependent mass correction, which plays an important role in the MSSM's light stop scenario \cite{Carena:1996wj}.  
We expect that this effect will weaken the PT strength on a parameter point-by-point basis without affecting our broader conclusions regarding testability of models with a first order electroweak phase transition.  

%==========
We assume that the new scalar fields do not acquire vevs, $\langle \tilde{u} \rangle = \langle \tilde{d} \rangle = \langle \tilde{U} \rangle = 0$.  
Thus, the the flavor symmetry prevents the Higgs from mixing with the squark-like fields.  
Nevertheless, the trilinear interactions allows the stop-like fields to mix after electroweak symmetry breaking.  
The mixing angle satisfies
\begin{align}
	\tan 2 \theta = \frac{\sqrt{2} a_{hQU} v}{m_Q^2 - m_U^2 + \frac{1}{2} ( \lambda_{hQ} + \lambda_{hQ}^{\prime} - \lambda_{hU} ) v^2}
	\per
\end{align}
The spectrum consists of two ``stops'' and one ``sbottom,'' and we denote their mass eigenvalues by $M_{\tilde{t}_1}$, $M_{\tilde{t}_2}$, and $M_{\tilde{b}}$.  
Their couplings to the Higgs are given by 
\begin{subequations}\label{eq:g_stop}
\begin{align}
	g_{h\tilde{t}_1\tilde{t}_1} &= -\cos^2\theta \, \bigl( \lambda_{hQ} + \lambda_{hQ}' \bigr)v \,-\, \sin^2 \theta \, \lambda_{hU} v \,+\, \frac{a_{hQU} \sin 2\theta}{\sqrt{2}} \\
	g_{h\tilde{t}_2\tilde{t}_2} & =-\sin^2\theta \, \bigl( \lambda_{hQ} + \lambda_{hQ}' \bigr)v \,-\, \cos^2\theta \, \lambda_{hU} v \,-\, \frac{a_{hQU} \sin 2\theta}{\sqrt{2}} \\
	g_{h\tilde{t}_1\tilde{t}_2} & = -\frac{\sin2\theta}{2} \, \bigl( \lambda_{hQ} + \lambda_{hQ}' \bigr)v \,+\, \frac{\sin 2\theta}{2} \, \lambda_{hU} v \,-\, \frac{a_{hQU} \cos2\theta}{\sqrt{2}} \\
	g_{h\tilde{b}\tilde{b}} & = - \bigl( \lambda_{hQ}+\lambda_{hQ}'' \bigr) v
	\per
\end{align}
\end{subequations}
These dimensionful couplings can be read off of \eref{eq:VQU} upon diagonalizing the stop mass matrix.  

%==========
Since the stop- and sbottom-like particles are not actually colored, they do not affect the Higgs coupling to gluons.  
However, the new charged scalar particles do increase the strength of the Higgs-photon coupling radiatively.  
This increases the Higgs diphoton decay rate \cite{Djouadi:1996pb,Carena:2011aa}
\begin{align}\label{eq:stoplike_Ghgg}
	\Gamma_{h \to \gamma \gamma} = \frac{1}{64\pi} \frac{\alpha^2 M_h^3}{16\pi^2} \Bigl| \bar{A}_W + \bar{A}_t + \bar{A}_{\tilde{t}} + \bar{A}_{\tilde{b}} \Bigr|^2
\end{align}
where $\alpha \simeq 1/137$ is the electromagnetic fine structure constant and 
\begin{subequations}\label{eq:stoplike_A}
\begin{align}
	\bar{A}_{W} & = \frac{g_{hWW}}{M_W^2} F_{1}\bigl( M_h^2 / 4 M_W^2 \bigr) \\
	\bar{A}_{t} & = 2 N_{c} Q_{t}^2 \, \frac{g_{htt}}{M_t} \, F_{1/2} \bigl( M_h^2 / 4 M_t^2 \bigr) \\
	\bar{A}_{\tilde{t}} & = \sum_{i=1}^{2}  \, n_{f} Q_{\tilde{t}}^2 \, \frac{g_{h \tilde{t}_i \tilde{t}_i}}{M_{\tilde{t}_i}^2} \, F_{0} \bigl( M_h^2 / 4 M_{\tilde{t}_i}^2 \bigr) \\
	\bar{A}_{\tilde{b}} & = n_{f} Q_{\tilde{b}}^2 \, \frac{g_{h \tilde{b} \tilde{b}}}{M_{\tilde{b}}^2} \, F_{0} \bigl( M_h^2 / 4 M_{\tilde{b}}^2 \bigr)
\end{align}
\end{subequations}
with $g_{hWW} = g^2v/2 = 2 M_W^2/v$ and $g_{htt} = y_{t}/\sqrt{2} = M_t/v$.  
We take the electromagnetic charges of the stop- and sbottom-like particles to be $Q_{\tilde{t}} = 2/3$ and $Q_{\tilde{b}} = -1/3$.  
The functions $F(\tau)$ are defined in \rref{Djouadi:1996pb}.  
To compare with the SM contribution, $(\Gamma_{h\to\gamma \gamma})_{\SM}$, we drop the $\bar{A}_{\tilde{t}}$ and $\bar{A}_{\tilde{b}}$ terms from \eref{eq:stoplike_Ghgg}.  

%==========
The new charged scalars also affect the Higgs coupling to Z-bosons.  
At one-loop order the dominant effect typically comes from a wavefunction renormalization on the Higgs leg.  
This graphs brings two factors of the quartic couplings $\lambda$, which appear in \eref{eq:VQU}.  
The other one-loop graphs, {\it i.e.} the vertex renormalization and the Z-boson wavefunction renormalization, are suppressed compared to the Higgs wavefunction renormalization by factors of $e/\lambda$ where $e$ is the electromagnetic coupling.  
As long as $\lambda \gtrsim e \sim 0.1$ the Higgs wavefunction renormalization is the dominant effect on the $hZZ$ coupling.  
Consequently the $hZZ$ coupling deviates from its SM value by~\cite{Fan:2014axa}
\begin{align}\label{eq:stoplike_dZh}
%	\delta Z_h = -n_{f} \sum_{i,j=1}^2\frac{|g_{h\tilde{t}_i\tilde{t}_j}|^2}{32 \pi^2}\ I_B(M_h^2;  M_{\tilde{t}_i}^2, M_{\tilde{t}_j}^2) - n_{f} \frac{|g_{h\tilde{b}\tilde{b}}|^2}{32 \pi^2}\ I_B(M_h^2;  M_{\tilde{b}}^2, M_{\tilde{b}}^2\ )
	\delta g_{hZZ} = n_{f} \sum_{i,j=1}^2\frac{|g_{h\tilde{t}_i\tilde{t}_j}|^2}{32 \pi^2}\ I_B(M_h^2;  M_{\tilde{t}_i}^2, M_{\tilde{t}_j}^2) + n_{f} \frac{|g_{h\tilde{b}\tilde{b}}|^2}{32 \pi^2}\ I_B(M_h^2;  M_{\tilde{b}}^2, M_{\tilde{b}}^2\ )
\end{align}
where the loop function is defined in \eref{eq:IB_def}.  

%----------------------------------------------------------------
% Heavy Fermions
%----------------------------------------------------------------
\subsection{Heavy Fermions}\label{sec:Heavy_Fermions}

%==========
In the previous sections we have discussed how new scalar states can lead to a first order electroweak phase transition.  
Here we investigate models in which the first order transition derives from new fermions.  
The key ingredient is that the fermions have a large coupling to the Higgs, which gives them a large mass during the electroweak phase transition.  
Prior to the phase transition the fermions were lighter, since the Higgs vev doesn't contribute to their mass.  
Since it is energetically unfavorable for the fermion masses to grow, the effective potential develops a barrier, which leads to a first order phase transition.  
We discuss two implementations of this idea.  

%----------------------------------------------------------------
% Heavy Chiral Fermions
%----------------------------------------------------------------
\subsubsection{Heavy Chiral Fermions}\label{sec:Chiral_Fermions}

%==========
Extend the SM to include the left-chiral fermions $\tilde{H}_1$, $\tilde{H}_2$, $\tilde{B}$, and $\tilde{W}^a$.  
They have the same gauge charge assignments as Higgsinos and gauginos in the MSSM.  
Two doublets $\tilde{H}_{1,2}$ are required for anomaly cancellation, and either the singlet $\tilde{B}$ or the triplet $\tilde{W}^{a}$ is required to allow a Yukawa coupling with the Standard Model Higgs.  
Working in the two-component Weyl spinor notation, the Lagrangian is
\begin{align}
	\Lcal & = \Lcal_{\SM} + \tilde{H}_1^{\dagger} i \bar{\sigma}^{\mu} D_{\mu} \tilde{H}_1 + \tilde{H}_2^{\dagger} i \bar{\sigma}^{\mu} D_{\mu} \tilde{H}_2 + \tilde{B}^{\dagger} i \bar{\sigma}^{\mu} \partial_{\mu} \tilde{B} + \tilde{W}^{a \dagger} i \bar{\sigma}^{\mu} \partial_{\mu} \tilde{W}^a \nn
	& \quad - \frac{1}{2} M_1 \bigl[ \tilde{B} \tilde{B} + \hc \bigr] - \frac{1}{2} M_2 \bigl[ \tilde{W}^a \tilde{W}^a + \hc \bigr] - \mu \bigl[ \tilde{H}_2 \cdot \tilde{H}_1 + \hc \bigr] \nn
	& \quad - \bigl[ \Phi^{\dagger} \bigl( h_2 \sigma_a \tilde{W}^a + h_2^{\prime} \tilde{B} \bigr) \tilde{H}_2 + \hc \bigr] - \bigl[ \Phi \cdot \bigl( -h_1 \sigma_a \tilde{W}^a + h_1^{\prime} \tilde{B} \bigr) \tilde{H}_1 + \hc \bigr] 
	\per
\end{align}
We have again used a SUSY-like notation for the fields so as to easily identify their gauge quantum numbers. 

%==========
Reference~\cite{Carena:2004ha} studied the phenomenology in this model and identified a region of parameter space in which the electroweak phase transition can be first order.  
In general, the model has $7$ free parameters:  $3$ mass parameters $\{ M_1, M_2, \mu \}$ and $4$ couplings $\{ h_1, h_2, h_1^{\prime}, h_2^{\prime} \}$.  
\rref{Carena:2004ha} suggests to focus on the restricted parameter space
\begin{align}\label{eq:chiral_fix_params}
	M_1 = M_2 = - \mu 
	\, , \qquad
	h_1 = h_2 \equiv h
	\, , \quad \text{and} \qquad
	h_1^{\prime} = h_2^{\prime} \equiv h^{\prime}
	\per
\end{align}
This reduces the free parameters to $3$: one mass parameter $\mu$ and two couplings $h$ and $h^{\prime}$.  
In this restricted parameter space, the spectrum consists of two degenerate ``charginos,'' two degenerate lighter ``neutralinos,'' and two degenerate heavier ``neutralinos'':  
\begin{subequations}\label{eq:fermion_spectrum}
\begin{align}
	M^2_{\tilde{C}_1} & = M^2_{\tilde{C}_2} = \mu^2 + h^2 v^2 \\
	M^2_{\tilde{N}_1} & = M^2_{\tilde{N}_2} = \mu^2 \\
	M^2_{\tilde{N}_3} & = M^2_{\tilde{N}_4} = \mu^2 + \bigl( h^2 + h^{\prime 2} \bigr) v^2 
	\per
\end{align}
\end{subequations}
The mass eigenstates have couplings with the Higgs given by 
\begin{subequations}\label{eq:fermion_g}
\begin{align}
	g_{h\tilde{N}_i\tilde{N}_j} & = \frac{1}{\sqrt{2}}\{ (h' N_{i1} - h N_{i2}) (N_{j3}-N_{j4}) + i \leftrightarrow j \} \\
	g_{h\tilde{C}_i\tilde{C}_j} & = \frac{1}{\sqrt{2}} h \delta_{ij}
	\com
\end{align}
\end{subequations}
where the neutralino $\tilde{N}_i$ can be decomposed into $N_{i1} \tilde{B} + N_{i2} \tilde{W} + N_{i3} \tilde{H}_2 + N_{i4} \tilde{H}_1$.  
We only consider $\mu > M_h/2$ to avoid the Higgs and Z invisible decays.

%==========
\rref{Carena:2004ha} identified that the phase transition can be strongly first order in the limit $\mu \ll h v, h'v$ with $h$ or $h' = O(1)$.  
In this case, the charginos and neutralinos are light near in the symmetric phase ($v \to 0$ in \eref{eq:fermion_spectrum}) but heavy in the broken phase.  
This makes it energetically preferable for the system to remain in the symmetric phase, and the phase transition is delayed until a lower temperature thereby becoming more strongly first order.  
Since we will be interested in a region of parameter space with large couplings, there is a threat that the new fermions will exacerbate the electroweak vacuum instability.  
Following \rref{Carena:2004ha} we counter this problem by introducing new scalar particles that have field-dependent masses $\mu_s^2 + (h^2 + h^{\prime 2}) \phi_h^2$ where $\mu_s^2 = {\rm exp}( 8 \pi^2 M_h^2 / (4 (h^2 + h^{\prime 2})^2 v^2) ) M_{\tilde{N}_{3}}^2 - (h^2 + h^{\prime 2}) v^2$.  
We do not expect these particles to play any significant role in the phenomenology or phase transition dynamics.  

%==========
The presence of new charged fermions tends to suppress the Higgs diphoton decay rate.  
We calculate $\Gamma_{h \to \gamma \gamma}$ by generalizing the MSSM chargino calculation in \rref{Djouadi:1996pb}:  
\begin{align}\label{eq:fermion_Ghgg}
	\Gamma_{h \to \gamma \gamma} = \frac{1}{64\pi} \frac{\alpha^2 M_h^3}{16\pi^2} \, \Bigl| \bar{A}_W + \bar{A}_t + \bar{A}_{\tilde{C}} \Bigr|^2
\end{align}
where $\bar{A}_{W}$ and $\bar{A}_{t}$ appear in \eref{eq:stoplike_A} and the chargino contribution is given by 
\begin{align}\label{eq:fermion_A}
	\bar{A}_{\tilde{C}} = \sum_{i=1}^{2} \, 2 \frac{g_{h\tilde{C}_i \tilde{C}_i}}{M_{\tilde{C}_i}} \, F_{1/2}\bigl(M_h^2/4 M_{\tilde{C}_i}^2 \bigr)
	\per
\end{align}
We will see that if $h = O(1)$ the model is already strongly constrained by LHC limits on the Higgs diphoton decay width.  
Therefore we also consider the case $h = 0$ where the charginos do not contribute to the Higgs diphoton decay rate at one-loop order.  

%==========
Both the charginos and neutralinos affect the Higgs coupling to Z-bosons.  
As argued above \eref{eq:stoplike_dZh}, the dominant effect comes from the Higgs wavefunction renormalization provided that $h,h^{\prime} \gtrsim e \sim 0.1$.  
Thus we calculate the deviation to the $hZZ$ coupling as 
\begin{align}\label{eq:fermion_dZh}
%	\delta Z_h = \sum_{i,j=1}^4\frac{ |g_{h\tilde{N}_i\tilde{N}_j}|^2 v^2}{32 \pi^2}\ I_F (M_h^2;  M_{\tilde{N}_i}^2, M_{\tilde{N}_j}^2) + 2\sum_{i,j=1}^{2} \frac{|g_{h\tilde{C}_i\tilde{C}_j}|^2v^2}{32\pi^2} \ I_F(M_h^2; M_{\tilde{C}_{i}}^2, M_{\tilde{C}_{j}}^2)
	\delta g_{hZZ} = - \sum_{i,j=1}^4\frac{ |g_{h\tilde{N}_i\tilde{N}_j}|^2 v^2}{32 \pi^2}\ I_F (M_h^2;  M_{\tilde{N}_i}^2, M_{\tilde{N}_j}^2) - 2\sum_{i,j=1}^{2} \frac{|g_{h\tilde{C}_i\tilde{C}_j}|^2v^2}{32\pi^2} \ I_F(M_h^2; M_{\tilde{C}_{i}}^2, M_{\tilde{C}_{j}}^2)
\end{align}
where the fermion loop integral is given by 
\begin{align}\label{eq:IF_def}
	I_F(p^2; m_1^2,m_2^2) = 4 \int_0^1 \! \ud x \ \Biggl( 
	& \frac{1}{6} 
	+ \frac{(1-x)x \bigl( m_1 m_2 - 2 m_1^2 x - 2 m_2^2 + p^2 (1-x)x \Bigr)}{m_2^2 (1-x) + m_1^2 x - p^2 (1-x)x} 
	\nn & 
	- (1-x) x \, \log \frac{m_2^2 (1-x) + m_1^2 x - p^2(1-x)x}{M_h^2} 
	\Biggr)
\end{align}
where $M_h^2$ appears as the $\overline{\rm MS}$ renormalization scale.  

%==========
Since the $hZZ$ coupling arises already at tree-level, $g_{hZZ} = (g^2 + g^{\prime 2}) v/2$, there can be another one-loop correction\footnote{We thank Michael Fedderke for bringing this issue to our attention.} when the $\overline{\rm MS}$ parameters $g$ and $g^{\prime}$ are matched onto physical quantities such as the $Z$ boson mass and $G_F$ \cite{Craig:2014una}. In this model, we expect a large contribution from the matching because the custodial symmetry is broken. 
Using Ref.~\cite{Craig:2014una} we estimate that this matching effect can contribute to $\delta g_{hZZ}$ with a magnitude that is comparable to \eref{eq:fermion_dZh}, but a more precise calculation of $\delta g_{hZZ}$ is beyond the scope of our work.  
Since we are primarily interested in establishing that models with a first order phase transition are testable at future colliders, we are satisfied that the predicted value of $\delta g_{hZZ}$ in this model is comparable to, or larger than,\footnote{We neglect the unnatural possibility that there is a tuning between these two independent contributions to $\delta g_{hZZ}$. } the value given by \eref{eq:fermion_dZh}.  

%==========
As discussed previously, the custodial symmetry is broken in this model, and therefore, we expect a large contribution to the electroweak precision parameter, T~\cite{Fedderke:2015txa}.  
To avoid the constraints, we can introduce new particles to the theory to cancel the contribution.  
A heavy Higgs can generate a negative contribution to the T parameter, and compensate the contribution to the T parameters from the neutralinos and the charginos.  

%----------------------------------------------------------------
% Varying Yukawa Couplings
%----------------------------------------------------------------
\subsubsection{Varying Yukawa Couplings}\label{sec:Vary_Yukawa}

%==========
References~\cite{Baldes:2016rqn,Baldes:2016gaf} recently studied the electroweak phase transition in an electroweak-scale implementation of the Froggatt-Nielsen mechanism.  
We investigate their model in this section.  

%==========
Extend the Standard Model to include a real scalar field $\chi$, which is a singlet under the Standard Model gauge group.  
We call $\chi$ the flavon field.  
By enforcing an appropriately chosen flavor symmetry, the Standard Model Yukawa interactions can be forbidden.  
Instead, the Yukawa interactions are generated from dimension-five operators after the flavon gets a vev.  
These dimension-five operators are written schematically as 
\begin{align}\label{eq:Lint_vary_Yuk}
	\Lcal_{\rm int} = - \left( \frac{\chi}{M} \right)^{q_i - q_{H} - q_j} \overline{f}_{L}^{i} \Phi f_{R}^{j}
\end{align}
where the $q$'s are flavor charges of the fermions and the Higgs.  
The associated Yukawa matrix is $y_{ij} = (\langle \chi \rangle / M)^{q_i-q_H-q_j}$.  

%==========
In \rref{Baldes:2016rqn} it is assumed that the expectation value of $\chi$ changes during the electroweak phase transition.  
(It need not be zero before the transition.)  
Across the two-dimensional field space, the fermion masses vary due to the explicit dependence on the Higgs field and the implicit dependence on the flavon field, via the Yukawa coupling.  
For fermion species $f \in \{ e, \mu, \tau, u, d, c, s, t, b\}$ we can write the field-dependent mass as 
\begin{align}\label{eq:vary_Yuk_mass}
	M_{f}(\phi_h,\chi) = \frac{y_{f}(\chi) \phi_h}{\sqrt{2}}
	\per
\end{align}
The electroweak phase transition can be strongly first order if $y_{f} \sim O(1)$ in the symmetric phase and $y_{f} \ll 1$ in the broken phase \cite{Baldes:2016rqn}.  
In this case the fermion is light in both the symmetric and broken phases, but heavy at intermediate field values.  
As a result, intermediate field values are energetically disfavored, and the corresponding potential energy barrier induces the phase transition to become first order.  
One can view this scenario as a different implementation of the heavy fermion model of \sref{sec:Chiral_Fermions}.  

%==========
One could study the electroweak phase transition in this model by specifying a potential for $\chi$ and tracking the evolution of both $\chi$ and the Higgs $\phi_h$ through the two-dimensional field space \cite{Baldes:2016gaf}.  
For simplicity, we follow \rref{Baldes:2016rqn} in assuming that $\chi$ can be parametrized in terms of $\phi_h$ along the phase transition trajectory.  
Then the model is specified by writing $y_f(\phi_h)$.  
Taking a phenomenological perspective, we consider three models for the Yukawa couplings 
\begin{align}\label{eq:Yuk_3models}
%%%
	(A) & \qquad y_f(\phi_h) = \begin{cases}
	y_1 \left( 1 - \frac{\phi_h^2}{v^2} \right) + \sqrt{2} M_f / v & \qquad \text{, for $f$ a quark} \\
	\sqrt{2} M_f / v & \qquad \text{, for $f$ a lepton} 
	\end{cases} \nn
%%%
	(B) & \qquad y_f(\phi_h) = \begin{cases}
	\sqrt{2} M_f / v & \qquad \text{, for $f$ a quark} \\
	y_1 \left( 1 - \frac{\phi_h^2}{v^2} \right) + \sqrt{2} M_f / v & \qquad \text{, for $f$ a lepton} 
	\end{cases} \\
%%%
	(C) & \qquad y_f(\phi_h) = \begin{cases}
	\sqrt{2} M_f / v & \qquad \text{, for $f$ a quark} \\
	\left[ 100 y_1 \left( 1 - \frac{\phi_h^2}{v^2} \right) + 1 \right] \times \sqrt{2} M_f / v & \qquad \text{, for $f$ a lepton} 
	\end{cases} \nonumber
\end{align}
where $M_f$ is the mass of fermion $f$.  
In each case, the Yukawa coupling is fixed to $\sqrt{2}M_f/v$ for $\phi_h \geq v$.  

%==========
The first two models are chosen to match the parametrization studied in \rref{Baldes:2016rqn}.  
Here, the varying flavon field leads to a universal (flavor-independent) {\it shift} in the Yukawa couplings for either the quarks or the leptons.  
The third model is chosen to avoid possible constraints from flavor-changing neutral currents.  
Here, the flavon leads to a flavor-independent {\it rescaling} of the lepton Yukawa couplings while leaving the quark Yukawa couplings unaffected.  
The factor of $100$ is introduced to balance the $\tau$ Yukawa coupling $\sqrt{2} M_{\tau} / v \simeq 0.01$; then, $y_{\tau}(0) \simeq y_1$ and $y_{e,\mu}(0) \simeq y_1 M_{e,\mu} / M_{\tau} \ll y_1$.  
Whereas the first two models can be thought to arise from the operator in \eref{eq:Lint_vary_Yuk}, the third model would arise from the operator $\Lcal_{\rm int} = - (\chi/M)^n y_{ij} \overline{f}_L^i \Phi f_{R}^j$ where vev of the flavon field sets the scale of the Yukawa matrix but not the flavor-dependent mixings and splittings.  

%==========
Since we have not specified an explicit interaction between the flavon and Standard Model fields, it is difficult to assess collider constraints on this model.  
In general, we expect that the model can be constrained via Higgs-flavon mixing and flavor-changing neutral currents (FCNC).  
Generally, the physical Higgs boson and the singlet flavon will mix, unless a symmetry forbids it.  
As we discussed in \sref{sec:Singlet}, this mixing leads to a number of potential constraints.  
Second, if the flavon's interactions are not flavor-diagonal, then it may contribute to FCNC, which are strongly constrained.  For instance, as discussed in~\cite{Baldes:2016gaf}, the scenario with one Froggatt-Nielsen flavon is ruled out, and the scenrio with two Froggatt-Nielsen flavons are constrained from Higgs and top exotic decays, and other flavor constraints.  

%==================================
% Probes of the Electroweak Phase Transition
%==================================
\section{Probes of the Electroweak Phase Transition}\label{sec:Probes}

%----------------------------------------------------------------
% Collider Probes
%----------------------------------------------------------------
\subsection{Collider Probes}\label{sec:HiggsFactory}

%==========
In the models we consider, the Higgs boson couples to new states that are typically below the TeV scale, and therefore, the models are subject to collider tests. 

%============
The new states we consider can contribute to the wavefunction renormalization of the Higgs.  
This affects the Higgs couplings universally.  
Among all the Higgs couplings, a lepton collider can measure the Higgs coupling to Z-bosons very well.  
This is accomplished in ``Higgs factory'' mode by running the lepton collider at $240 \GeV$ to $250 \GeV$ where the Higgs-strahlung cross section is maximized and the threshold for $l^+l^- \to Z^{\ast} \to hZ$ just opens up.  
Deviations in the $hZZ$ coupling from the SM expectation are parametrized by $\delta g_{hZZ}$, which was defined in \eref{eq:dZh_def}.  
The current LHC limit \cite{ATLAS-CONF-2015-044} and improved sensitivities with future colliders are summarized below:  
\begin{align}\label{eq:dZh_sensitivity}
	\Delta \bigl( \delta g_{hZZ} \bigr) = 
	\ & 27\% \ \text{(current)}
	\, , \ \ 
	7\% \ \text{(HL-LHC)}
	\, , \ \ 
	0.25\% \ \text{(CEPC, ILC-500)}
	\, , \ \ 
	0.15\% \ \text{(FCC-ee)}
	\per
\end{align}
Here we show sensitivities for the high-luminosity Large Hadron Collider (HL-LHC) \cite{CMS:2013xfa}, Circular Electron Positron Collider (CEPC) \cite{CEPC-SPPCStudyGroup:2015csa}, International Linear Collider at $\sqrt{s} = 500 \GeV$ (ILC-500) \cite{Fujii:2015jha}, and Future Circular Collider study with both $240$ and $350 \GeV$ measurements (FCC-ee) \cite{dEnterria:2016cpw}.  

%============
New charged states coupled to the Higgs will radiatively affect the Higgs decay rate into a pair of photons.  
Both current and future colliders can constrain a deviation from the SM prediction in this channel. 
We summarize below the current sensitivity of the ATLAS experiment \cite{ATLAS-CONF-2015-044}, the projected sensitivity of the CMS experiment at the HL-LHC \cite{CMS:2013xfa}, and the expected sensitivity of future lepton colliders \cite{CEPC-SPPCStudyGroup:2015csa,dEnterria:2016cpw}:  
\begin{align}\label{eq:Higgs_diphoton_sensitivity}
	\Delta \Bigl( \frac{\Gamma_{h\to \gamma \gamma}}{(\Gamma_{h\to \gamma \gamma})_{\SM}} \Bigr)
	= 20\% \ \text{(current)}
	\ , \quad
	8\% \ \text{(HL-LHC)}
	\ , \quad
	4\% \ \text{(CEPC)}
	\ , \quad
	1.5\% \ \text{(FCC-ee)}
	\per
\end{align}

%============
The cubic self-coupling of the Higgs boson $\lambda_3$ is currently unconstrained by the LHC.  
The $500 \GeV$ and the $1 \TeV$ options for ILC can also measure $\lambda_3$ from the Higgs-strahlung, WW fusion, and the ZZ fusion processes~\cite{Fujii:2015jha}.  
Future hadron colliders with higher center of mass energy, for example, SppC and FCC-hh, are expected to increase the sensitivity in $\lambda_3$ significantly \cite{Yao:2013ika,Barr:2014sga,Azatov:2015oxa,He:2015spf,Huang:2015tdv}, to the ball park of $10 \%$.  
At the same time, much more detailed studies are needed to produce more solid estimate by carefully taking into all possible channels and systematics. 
These sensitivities are summarized as follows: 
\begin{align}\label{eq:Higgs_cubic_sensitivity}
	\Delta \Bigl( \frac{\lambda_{3}}{\lambda_{3,\SM}} \Bigr)
	= 27\% \ \text{(ILC-500)}
	\ , \quad
	10\% \ \text{(SppC / FCC-hh / ILC-1000)}
	\per
\end{align}

%----------------------------------------------------------------
% Strongly First Order Phase Transition and Baryogenesis
%----------------------------------------------------------------
\subsection{Strongly First Order Phase Transition and Baryogenesis}\label{sec:EWBG}

%============
The matter / anti-matter asymmetry of the universe may have been generated at the electroweak phase transition; this scenario is known as electroweak baryogenesis.  
(For a review, see \rref{Riotto:1999yt}.)  
In this framework, the electroweak phase transition needs to be first order, such that phase coexistence occurs during the transition.  
At the boundary between phases, {\it i.e.} the bubble wall, the properties of particles and their interactions can vary rapidly.  
Specifically, thermal diffusion of $\SU{2}_L$ Chern-Simons number leads to efficient violation of baryon number outside of the bubbles (symmetric phase), but inside of the bubbles the weak gauge fields become massive, $M_W(T) \approx g v(T)/2$, and these processes (electroweak sphalerons) acquire a Boltzmann suppression.  
Thus, a baryon asymmetry can be generated outside of the bubble and diffuse into the bubble, where it is approximately conserved.  
To avoid washout of the baryon asymmetry in the Higgs phase, the electroweak order parameter must satisfy \cite{Shaposhnikov:1987tw} (see also \cite{Ahriche:2007jp, Fuyuto:2014yia})
\begin{align}\label{eq:washout_crit}
	\frac{v(T)}{T} \Bigr|_{T=T_{\PT}} \gtrsim 1.3
\end{align}
at the temperature of the phase transition, $T_{\PT}$.  
If \eref{eq:washout_crit} is satisfied, the phase transition is said to be ``strongly'' first order.  
For our numerical results, we evaluate the washout criterion at the bubble nucleation temperature $T_n$; see \aref{app:PT_calc}.  

%============
Although \eref{eq:washout_crit} is a necessary condition for electroweak baryogenesis, it is not a sufficient condition.  
Successful generation of the baryon asymmetry also requires a source of $CP$-violation, which is model-dependent.  
Additionally it requires that the bubble walls are not expanding too quickly \cite{Kozaczuk:2015owa}.  
These details are beyond the scope of our work.  
Therefore, we do not consider electroweak baryogenesis explicitly, and we do not calculate the relic baryon asymmetry.  
Rather, we assess whether the phase transition is strongly first order depending on whether it satisfies \eref{eq:washout_crit}; if the phase transition is not strongly first order, electroweak baryogenesis is not viable.  

%----------------------------------------------------------------
% Gravitational Waves
%----------------------------------------------------------------
\subsection{Gravitational Waves}\label{sec:Grav_Waves}

%============
A first order cosmological phase transition is expected to generate a stochastic background of gravitational waves \cite{Hogan:1986qda, Kamionkowski:1993fg}.  
In general, the gravitational waves arise from several sources.  
When bubbles of the Higgs phase meet one another and collide, the localized energy density generates a quadrupole contribution to the stress-energy tensor, which sources gravitational waves.  
Additionally, the passage of bubbles through the plasma creates magnetohydrodynamic turbulence and sound waves.  
Decay of the turbulence and damping of the sound waves can continue for multiple Hubble times, even after the phase transition is completed, providing additional sources of gravitational waves.  

%============
The spectrum of stochastic gravitational waves is very model-dependent.  
(We provide additional details in \aref{app:GW_calc}.)  
A key parameter is the ratio $\alpha = \Delta \rho_{\rm vac} / \rho_{\rm rad}$, which compares the vacuum energy density liberated in the phase transition with the energy density of the relativistic plasma.   
The gravitational wave spectrum is proportional to $\alpha^2$, up to some efficiency factor.  
Writing the phase transition temperature as $T_{\PT}$ we have $\alpha^2 \propto \rho_{\rm rad}^{-2} \propto T_{\PT}^{-8}$.  
In the Standard Model $T_{\PT} \simeq 160 \GeV$, but a first order phase transition (due to BSM physics) can have appreciable supercooling, typically $T_{\PT} \gtrsim 50 \GeV$.  
Thus the amplitude of the stochastic gravitational wave spectrum varies by many orders of magnitude across the parameter space.  

%============
Conversely, the peak frequency of the gravitational wave spectrum is less model-sensitive.  
The frequency of these gravitational waves today is related to the size of the Higgs phase bubbles at the time of collision.  
Typically, this is some (model-dependent) fraction of the cosmological horizon at the time of the phase transition.  
(The complete formulas appear in \aref{app:GW_calc}.)
For an electroweak-scale phase transition, the spectrum typically peaks in the range $f \sim (10^{-4} - 10^{-2}) \Hz$.  
Such a signal is potentially within reach of future space-based gravitational wave interferometers, such as eLISA.  

%============
The geometry of eLISA has not been finalized, and a number of designs are under investigation.  
Configurations differ in regard to number of links (4 or 6), arm length (1, 2, or 5 million kilometers), duration (2 or 6 years), and noise level (comparable to Pathfinder or greater).  
The corresponding sensitivities to a stochastic gravitational wave background are calculated in \rref{Caprini:2015zlo} for four configurations.  
In its more optimistic configurations, the sensitivity of eLISA could reach 
\begin{align}
	\Delta( \Omega_{\rm gw} h^2 ) \sim 10^{-14} \ \ \text{to} \ \ 10^{-12} \quad \text{at} \quad f = 5 \times 10^{-3} \Hz
	\per
\end{align}
In the next section, we compare the predicted gravitational wave signals in our models against these projected sensitivities.  

%==================================
% Results
%==================================
\section{Results}\label{sec:Results}

%============
For each of the models we scan approximately $5000$ points in the parameter space.  
The parameter ranges vary from model to model, but unless otherwise specified we generally allow the dimensionless couplings to vary between $-2$ and $2$.  
At each parameter point, we calculate the relevant collider observables ($\delta g_{hZZ}$, $\delta\Gamma_{h\gamma \gamma}$, $\lambda_3$), the phase transition order parameter ($v/T$), and the gravitational wave spectrum.  
Throughout this section, the scatter plots are color-coded:  an orange point indicates a first order phase transition, a blue point indicates a strongly first order phase transition (\ref{eq:washout_crit}), and a green point indicates a very-strong first order phase transition with potentially detectable gravitational wave signal at eLISA.

%----------------------------------------------------------------
% Real Scalar Singlet
%----------------------------------------------------------------
\subsection*{Real Scalar Singlet}

\begin{figure}[t]
\begin{center}
\includegraphics[height=6.0cm]{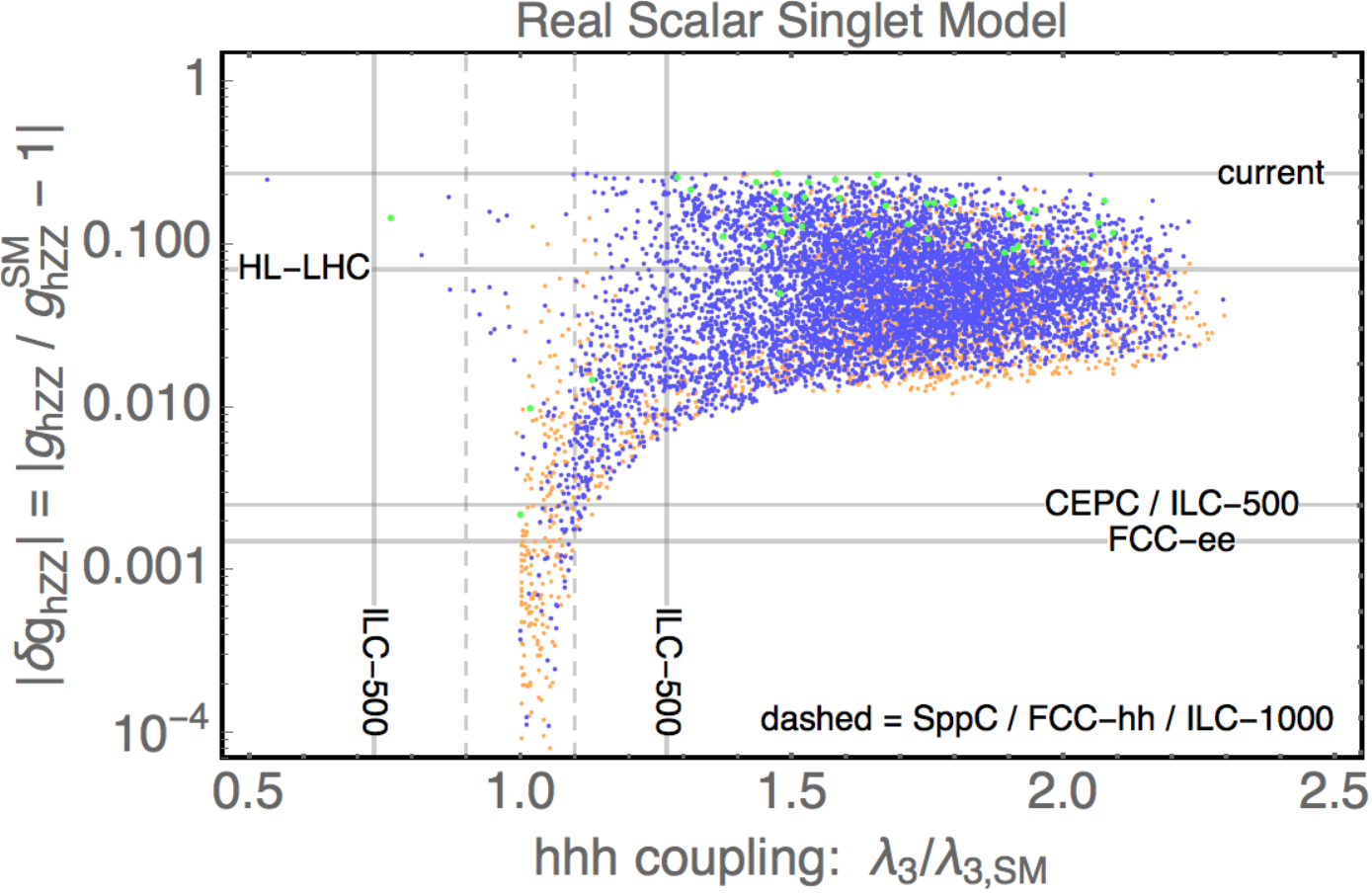} \hfill
\includegraphics[height=6.0cm]{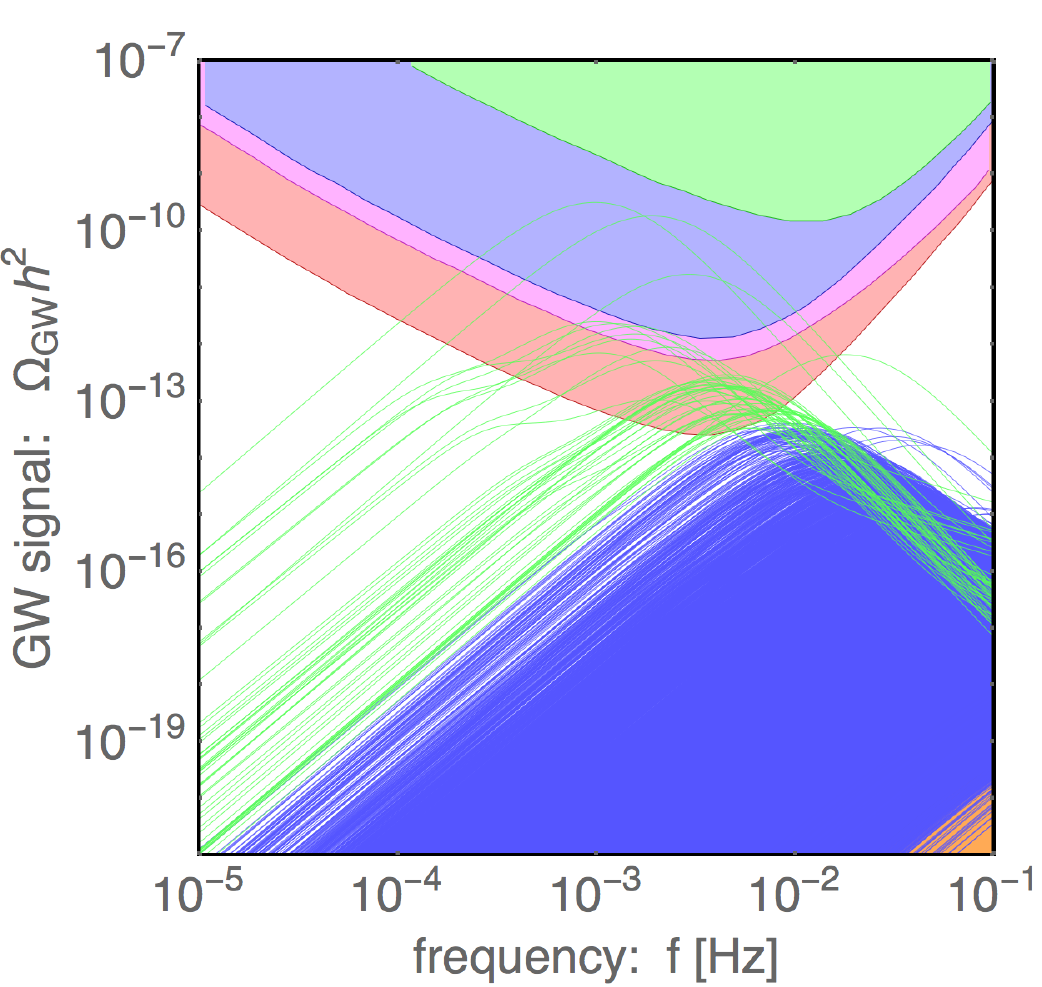}
\caption{
\label{fig:singlet_general_observables}
Parameter space scan for the singlet model of \sref{sec:Singlet}.  An orange point indicates a first order phase transition, a blue point indicates a strongly first order phase transition (\ref{eq:washout_crit}), and a green point indicates a very-strong first order phase transition with potentially detectable gravitational wave signal at eLISA. 
The right panels shows the predicted gravitational wave spectrum today along with the projected sensitivity of eLISA \cite{Caprini:2015zlo}.  
}
\end{center}
\end{figure}

%============
In \sref{sec:Singlet} we extended the Standard Model by a real scalar singlet, which is able to mix with the Standard Model Higgs.  
This mixing (in addition to radiative effects) leads to a modification of the $hZZ$ and $hhh$ couplings, parametrized by $\delta g_{hZZ}$ and $\lambda_{3}$.  
In \fref{fig:singlet_general_observables} we show the distribution of models over this parameter space.  
Models with a first order phase transition (orange, blue, or green points) tend to predict a large suppression of the $hZZ$ coupling, ranging from $\sim1\%$ to as much as $30\%$.  
For the vast majority of the models in the scan, this suppression is detectable at Higgs factories such as the CEPC, which has an expected sensitivity of $\Delta(\delta g_{hZZ}) = 0.25\%$ (\ref{eq:dZh_sensitivity}).  
Focusing on the models with a gravitational wave signal that is  potentially detectable by eLISA (green points), the suppression of $hZZ$ is larger still, typically $\gtrsim 10\%$.  
Therefore if the Higgs factory measures a significant suppression of the $hZZ$ coupling, it would motivate a search for relic gravitational waves with space based interferometers.   

%============

\begin{figure}[t]
\begin{center}
\includegraphics[height=6.0cm]{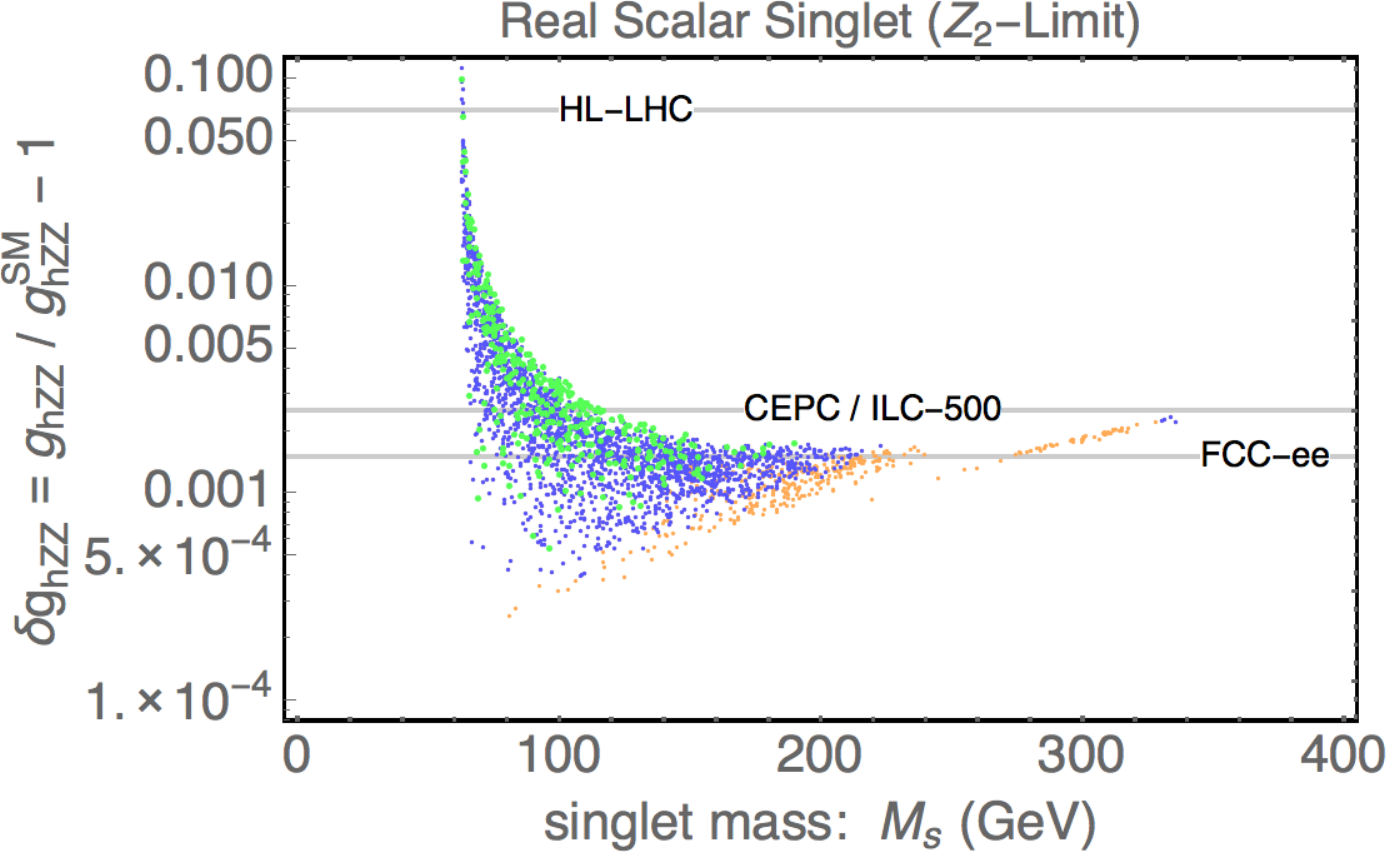} \hfill
\includegraphics[height=6.0cm]{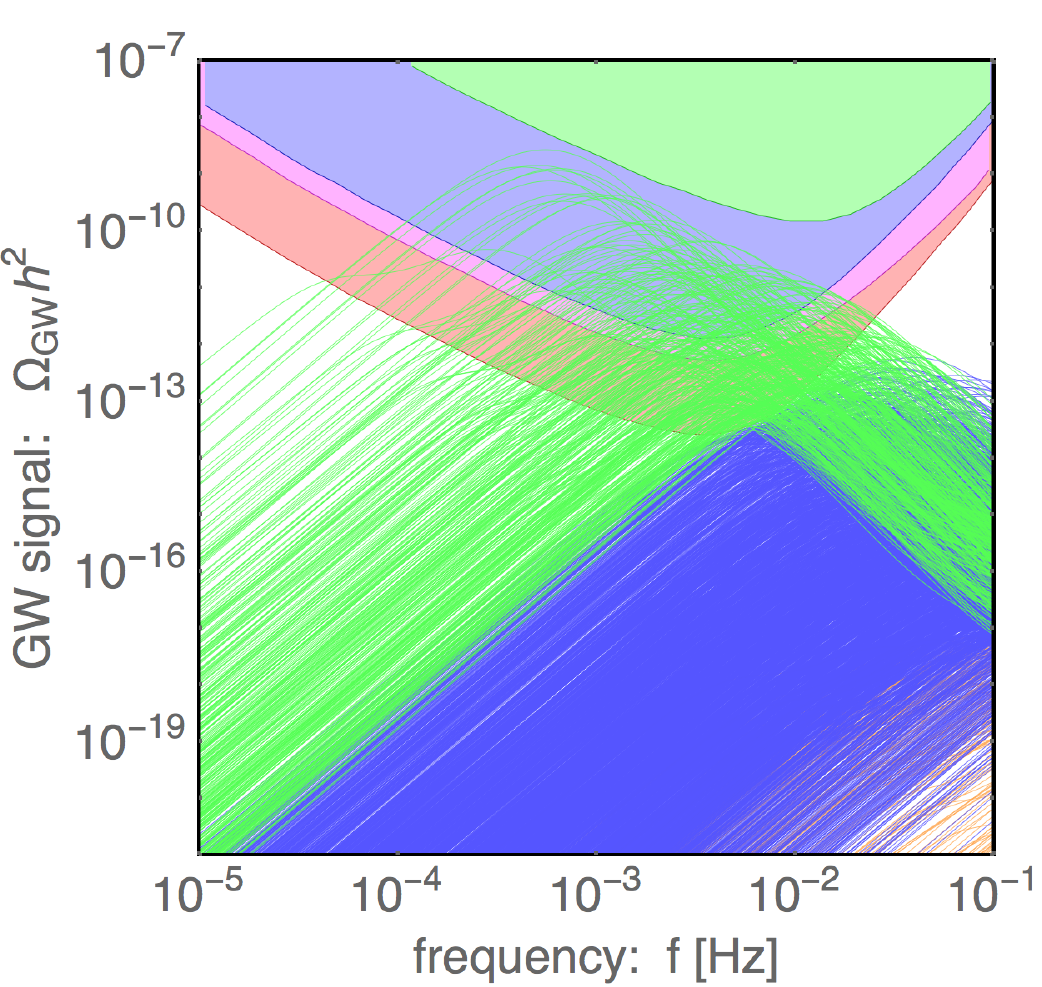} 
\caption{
\label{fig:singlet_Z2_observables}
Parameter space scan for the singlet model of \sref{sec:Z2} where a discrete $\Zbb_2$ symmetry forbids the Higgs-singlet mixing and suppresses the BSM modification to the $hZZ$ coupling.  
}
\end{center}
\end{figure}

%============
In the $\mathbb{Z}_2$-symmetric limit of \sref{sec:Z2}, the only free parameters are $M_s$, $\lambda_{hs}$, and $\lambda_{s}$.  
The discrete symmetry forbids a mixing between the Higgs and singlet fields, and deviations in the $hZZ$ and $hhh$ couplings arise first at one-loop order.  
The loop-induced contribution to $\delta g_{hZZ}$ typically falls below the projected sensitivity of future Higgs factories.  
The region with a viable first order phase transition are shown in \fref{fig:singlet_Z2_observables}.  
This limit of the singlet model admits strongly first order, two-step phase transitions in which the singlet field acquires a vev prior to electroweak symmetry breaking.  
The density of very strong phase transitions (green points) is higher, in part, because of a sampling effect; here we scan $3$ model parameters whereas we scan $5$ in \fref{fig:singlet_general_observables}.  
Our results broadly agree with more detailed analyses in the literature, {\it e.g.} \rref{Curtin:2014jma}.  

%============

\begin{figure}[t]
\begin{center}
\includegraphics[height=6.0cm]{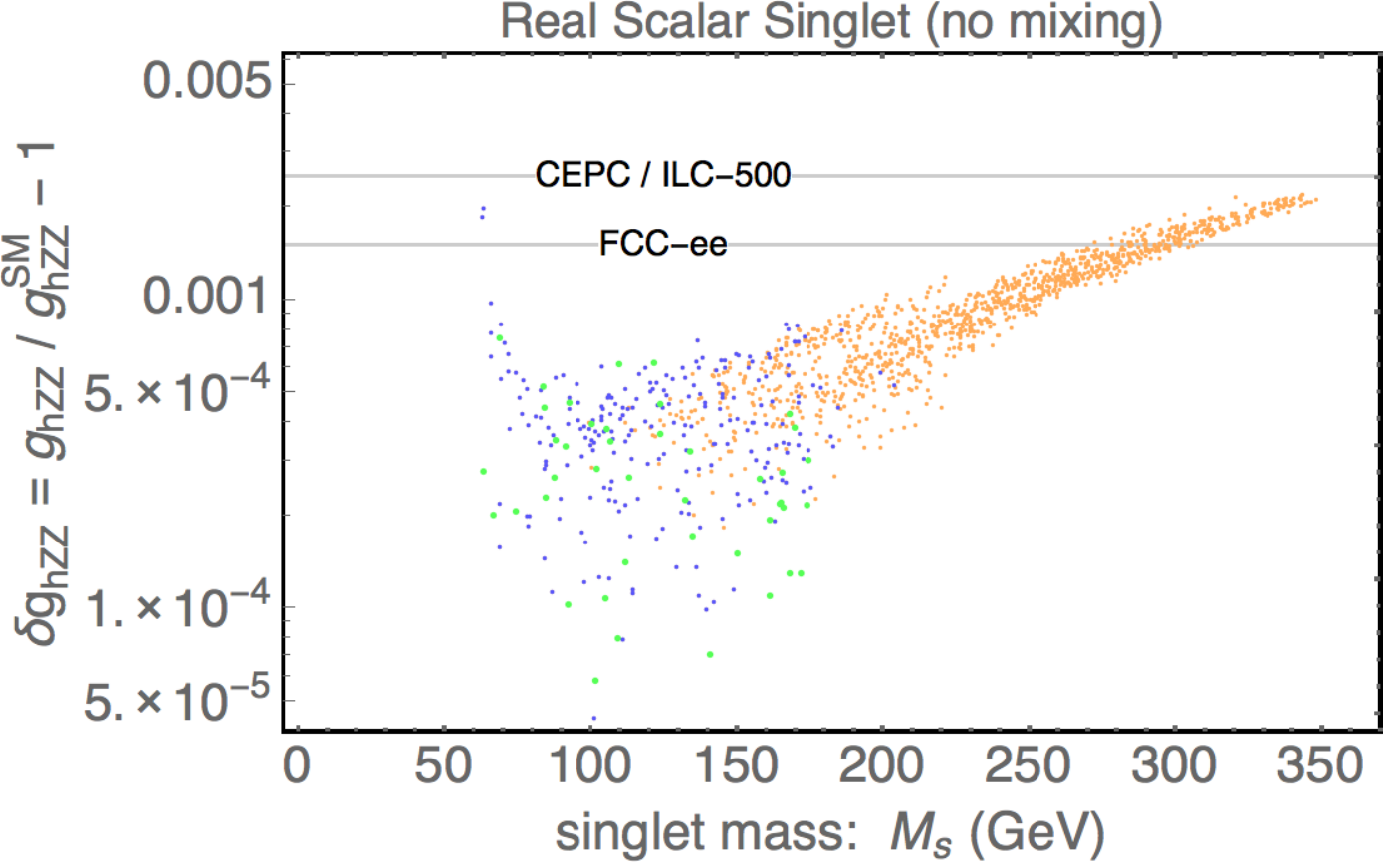} \hfill
\includegraphics[height=6.0cm]{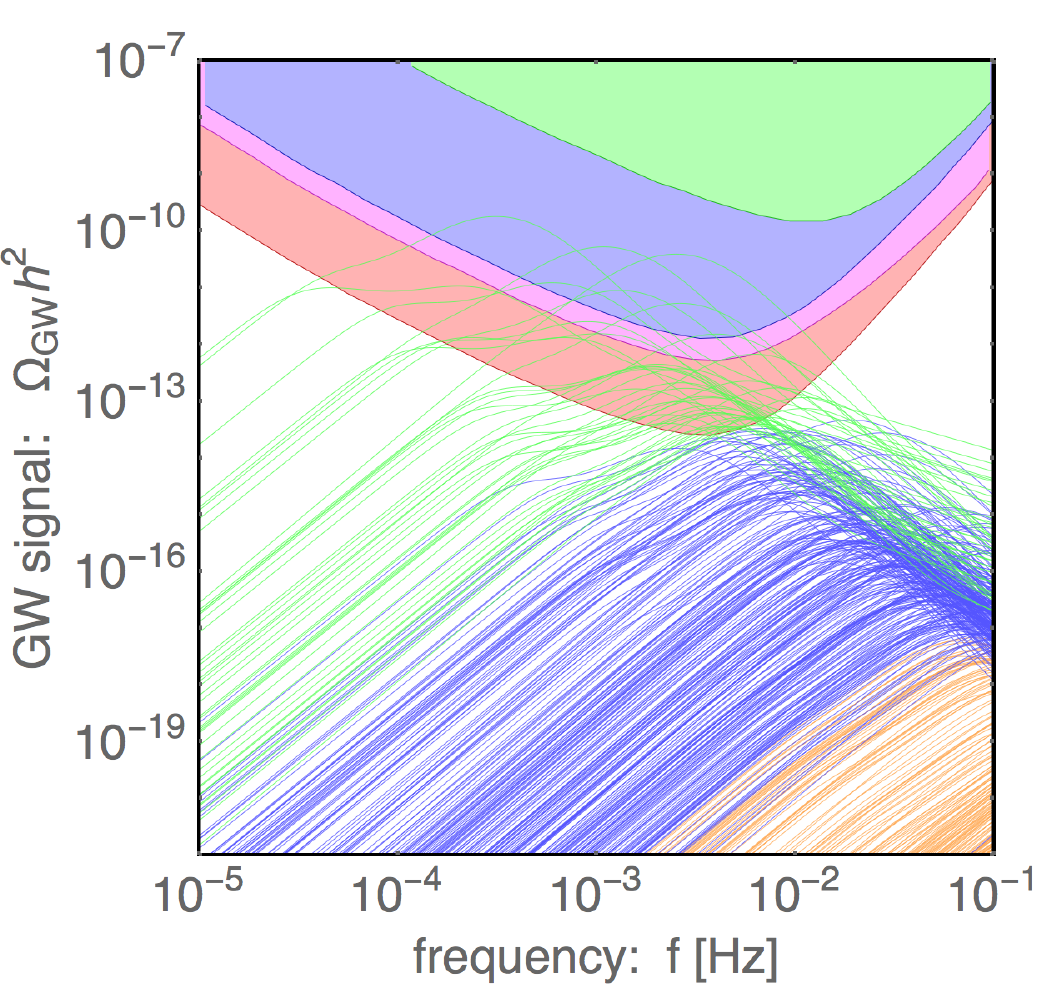}
\caption{
\label{fig:singlet_nomix_observables}
Parameter space scan for the singlet model of \sref{sec:Unmixed} where the Higgs-singlet mixing it tuned to zero.  
}
\end{center}
\end{figure}

%============
In \sref{sec:Unmixed} we discussed a limit of the scalar singlet model in which the mixing is tuned to zero.  
Then the $hZZ$ coupling is only modified at one-loop order, as seen in \eref{eq:singlet_dZh}.  
This region of parameter space is identified on \fref{fig:singlet_general_observables} as the ``funnel'' region where $\lambda_{3} / \lambda_{3,\SM} \to 1$ and $\delta g_{hZZ} \to 0$.  
We have performed a parameter scan focusing on this region, and we show the results in \fref{fig:singlet_nomix_observables}. 
Most of the parameter space corresponds to a weakly first order phase transition (orange points).  
In this limit, the phase transition occurs in two steps with the singlet first acquiring a negative vev at $T \gtrsim 200 \GeV$, and the electroweak symmetry is broken later $T \lesssim 100 \GeV$ when the singlet vev becomes positive.  
Due to the large field excursion and the barrier provided by tree-level potential terms ($a_{hs} \phi_h^2 \phi_s$ and $\lambda_{hs} \phi_h^2 \phi_s^2$) there is a significant amount of supercooling, and the phase transition is very strongly first order.  
However, at the zero-temperature vacuum, the model is very SM-like, and the deviation in the $hZZ$ coupling is too small to probe with future Higgs factories.  
Since the model admits strongly first order phase transition, but is inaccessible to collider probes, this limit can be viewed as a new class of ``nightmare scenario.''  

%============

%----------------------------------------------------------------
% Stop-Like Scenario
%----------------------------------------------------------------
\subsection*{Stop-Like Scenario}

\begin{figure}[t]
\begin{center}
\includegraphics[height=6.0cm]{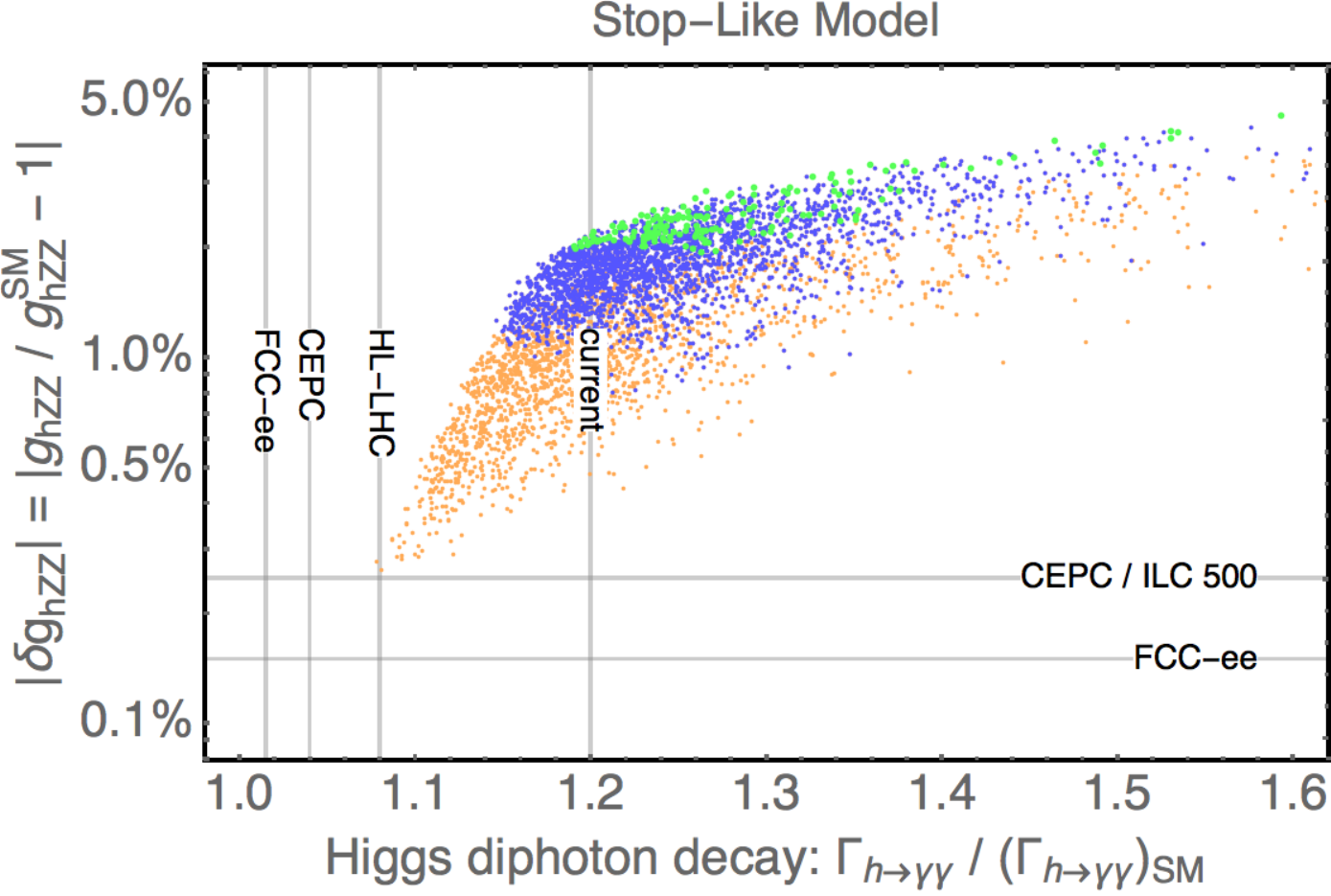} \hfill
\includegraphics[height=6.0cm]{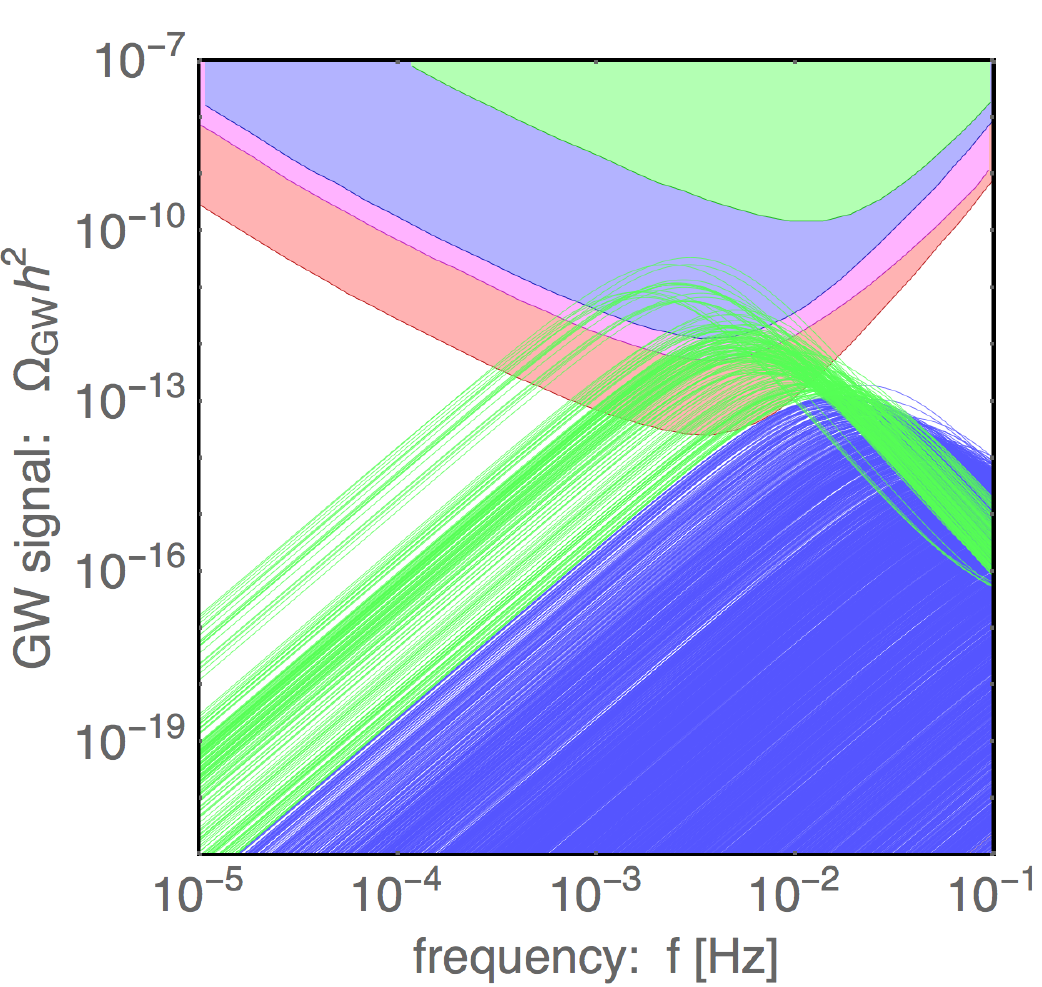} 
\caption{
\label{fig:stoplike_observables}
Parameter space scan for the stop-like model of \sref{sec:Stop_Like}.  The projected sensitivity of figure Higgs factories (CEPC, ILC-500, FCC-ee) is sufficient to test the entire region of parameter space where the phase transition is first order (orange, blue, \& green points).  
}
\end{center}
\end{figure}

%============
In \sref{sec:Stop_Like} we extend the SM by three scalar doublets and complex scalar singlets, which can be viewed as colorless stops and sbottoms.  
As the text discusses below \eref{eq:VQU}, we restrict to a 4-dimensional parameter space by assuming a common quartic coupling $\lambda$.  
The new charged scalars contribute to the Higgs diphoton decay width $\Gamma_{h\to \gamma \gamma}$ and lead to a deviation in the $hZZ$ coupling, parametrized by $\delta g_{hZZ}$.  
Figure~\ref{fig:stoplike_observables} shows the result of a scan over the 4-dimensional parameter space.  
In the region of parameter space with a first order phase transition (orange, blue, and green points), the Higgs diphoton decay width is enhanced by more than $10\%$, and it is enhanced by more than $20\%$ in the region with a potentially detectable gravitational wave signal (green).  
Given current LHC limits (\ref{eq:Higgs_diphoton_sensitivity}) some of this parameter space is already at tension with the data.  
More importantly, the projected sensitivity of figure Higgs factories (CEPC, ILC-500, FCC-ee) is sufficient to test the entire region of parameter space where the phase transition is first order.  

%============

%----------------------------------------------------------------
% Heavy Chiral Fermions
%----------------------------------------------------------------
\subsection*{Heavy Chiral Fermions}

%============
In \sref{sec:Chiral_Fermions} we extend the SM by four chiral fermions:  a pair of doublets (``Higgsinos''), a triplet (``wino''), and a singlet (``bino'').  
We vary the mass parameter $\mu$, the wino Yukawa coupling $h$, and the bino Yukawa coupling $h^{\prime}$.  
For simplicity we focus on the two cases, $h^{\prime} = 0$, $h \neq 0$ and $h=0$, $h^{\prime} \neq 0$.  

\begin{figure}[t]
\begin{center}
\includegraphics[height=5.5cm]{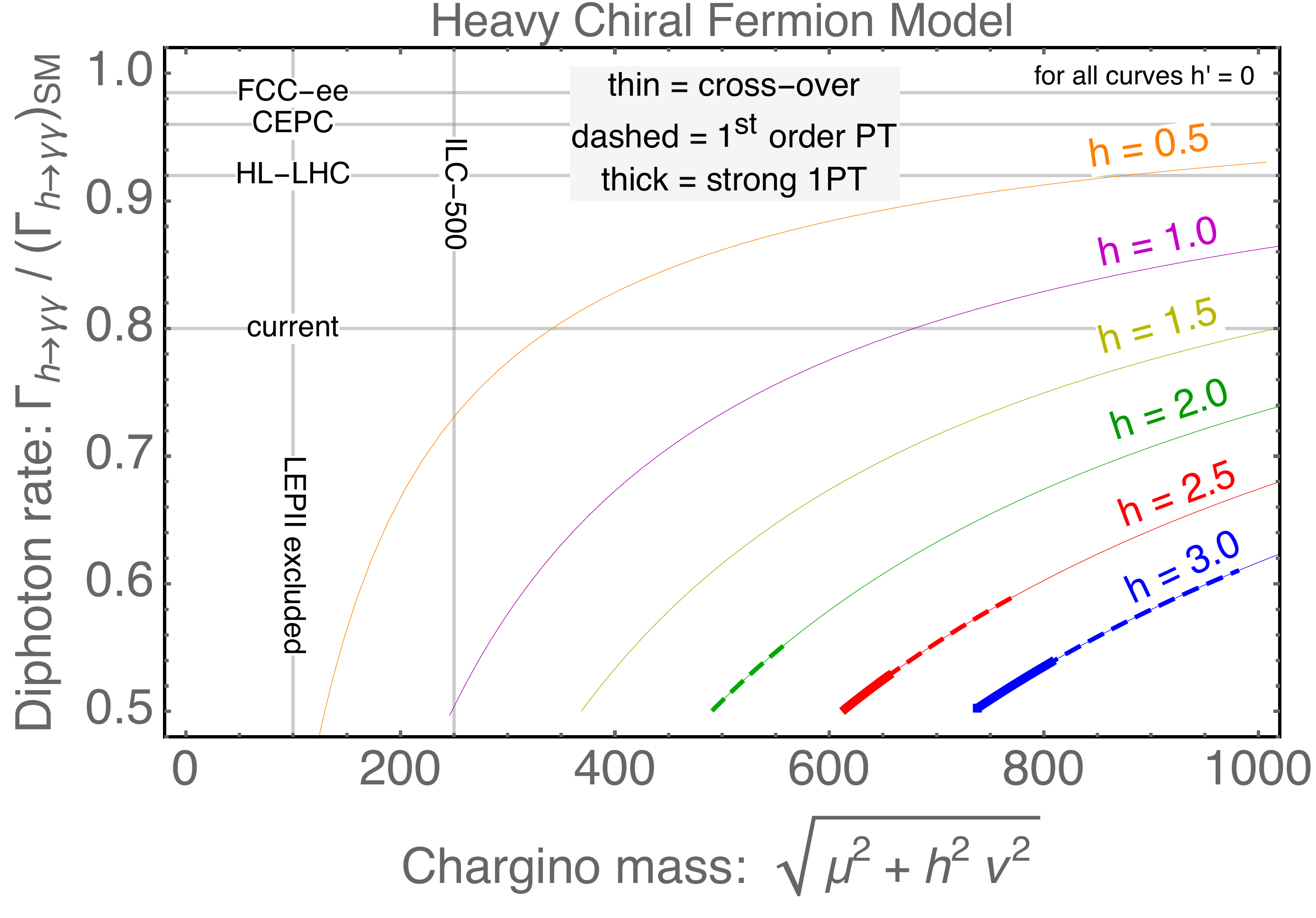} 
\caption{
\label{fig:heavy_fermion}
Results for the heavy chiral fermion model of \sref{sec:Heavy_Fermions}.  
The Higgs-to-diphoton decay rate is shown as a function of the chargino mass.  We fix the Higgs-bino-Higgsino Yukawa coupling $h^{\prime} = 0$ and we show various values of the Higgs-wino-Higgsino Yukawa coupling $h$.  Thick lines indicates parameters with a strongly first order phase transition (\ref{eq:washout_crit}), dashed lines indicate a weakly first order transition, and thin lines indicate a cross-over or second order transition.  
}
\end{center}
\end{figure}

%============
Nonzero $h$ allows the charginos to suppress the Higgs decay rate into photons (\ref{eq:fermion_Ghgg}).  
As seen in the left panel of \fref{fig:heavy_fermion} for $h = O(1)$ and $h^{\prime} = 0$ this suppression is already at tension with current limits on $\Gamma_{h \to \gamma \gamma}$ from the LHC (\ref{eq:Higgs_diphoton_sensitivity}).  
A large Yukawa coupling $h \gtrsim 2$ is required to achieve a first order phase transition (indicated by the dashed curves), and an even larger coupling $h \gtrsim 2.5$ is required for a strongly first order phase transition (thick solid curve).  
Although the phase transition is strongly first order in this region, the electroweak order parameter only gets as large as $v(T)/T \simeq 2.1$, and the corresponding gravitational wave signal is not within reach of eLISA.  

%============
Taking instead the wino Yukawa coupling to vanish ($h=0$) removes the tree-level interaction between Higgs and charginos, and the constraint from Higgs diphoton decay is avoided.  
We find that a first order phase transition is possible provided that the coupling is sufficiently large, $h^{\prime} \gtrsim 2$.  
However, this model is ruled out by a large deviation in the Peskin-Takeuchi T-parameter.  
As we discuss at the end of \eref{sec:Chiral_Fermions}, it may be possible to avoid constraints on $T$ by adding new particles to restore the custodial $\SU{2}$ symmetry.  
Then the model can still be tested through precision measurements of the $hZZ$ coupling.  
Using \eref{eq:fermion_dZh} we estimate the modification to $hZZ$ to be at the level of $O(10\%)$ if $h=0$ and $h^{\prime} = O(1)$.  
Such a large deviation can potentially be probed by the high-luminosity LHC or certainly by a future Higgs factory (\ref{eq:dZh_sensitivity}).  

%============
Finally let us comment on the size of the Yukawa couplings considered here.  
If $h$ and $h^{\prime}$ are as large as $2$ or $3$, as shown in \fref{fig:heavy_fermion}, then the model has a Landau pole not far above the electroweak scale.  
In principle new particles can be added to cancel the radiative corrections from the heavy chiral fermions and raise the cutoff.  
The new particles may enter Higgs physics, possibly making them testable at the LHC, but certainly testable at a future $pp$ collider.  

%============

%----------------------------------------------------------------
% Varying Yukawa Couplings
%----------------------------------------------------------------
\subsection*{Varying Yukawa Couplings}

%============
In \sref{sec:Vary_Yukawa} we allowed the Yukawa couplings of the Standard Model fermions to depend on the vacuum expectation value of a scalar flavon field.  
Provided that the flavon vev changes during the electroweak phase transition, the Yukawa couplings acquire an implicit dependence on the vev of the Higgs field.  
In \eref{eq:Yuk_3models} we discussed three phenomenological models:  (A) quark Yukawa couplings receive a universal (flavor-independent) {\it shift} and lepton Yukawa couplings are fixed, (B) lepton couplings are shifted and quark couplings are fixed, and (C) lepton couplings receive a universal {\it rescaling} and quark couplings are fixed.  

\begin{figure}[p]
\begin{center}
\includegraphics[height=6.0cm]{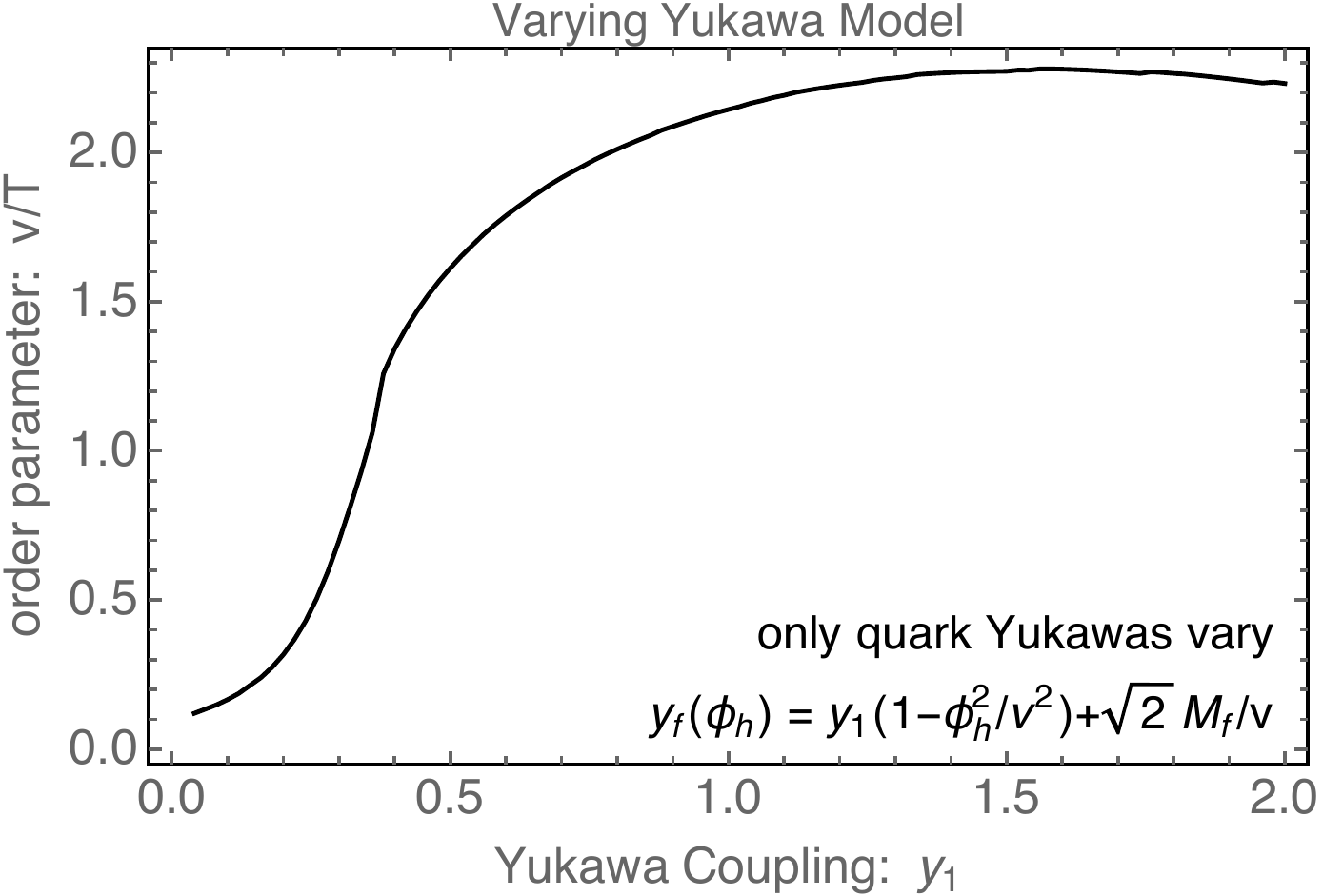} \hfill
\includegraphics[height=6.0cm]{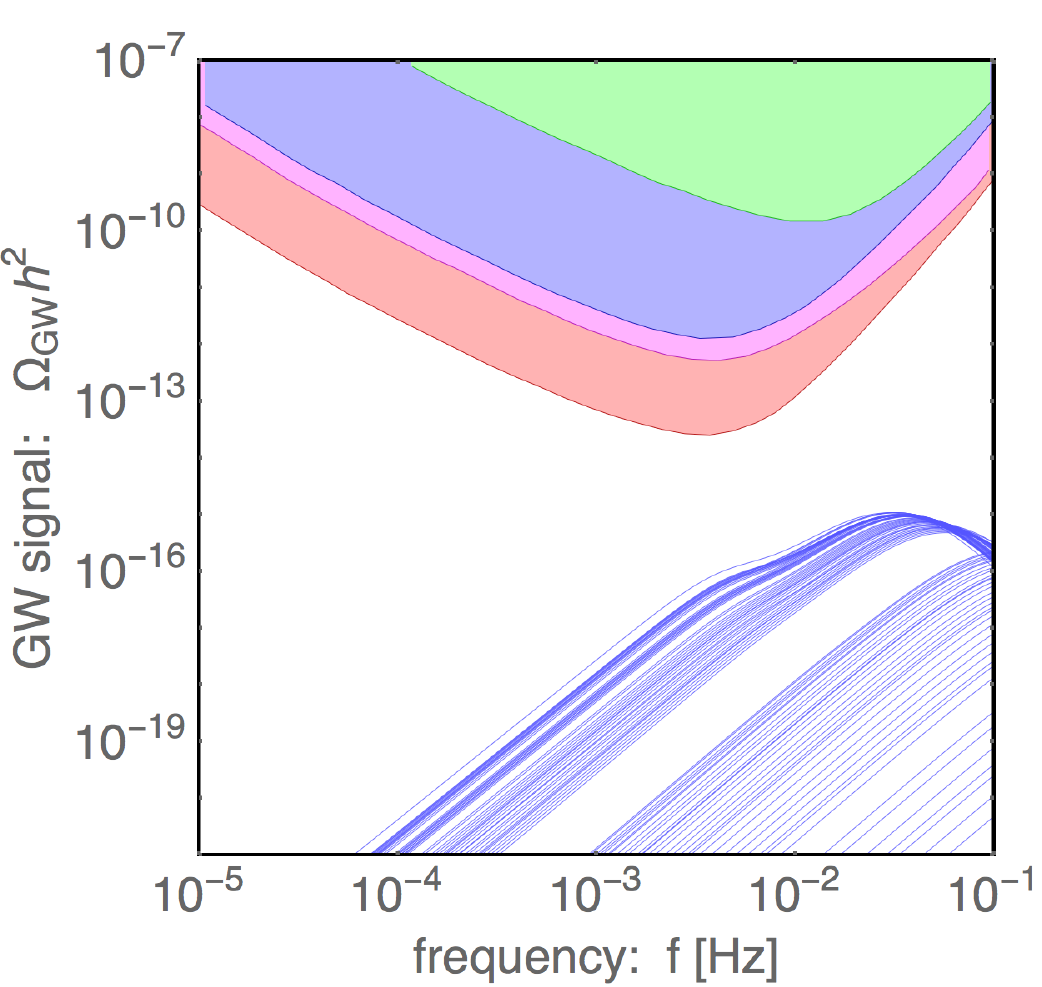} \\ \vspace{0.5cm}
\includegraphics[height=6.0cm]{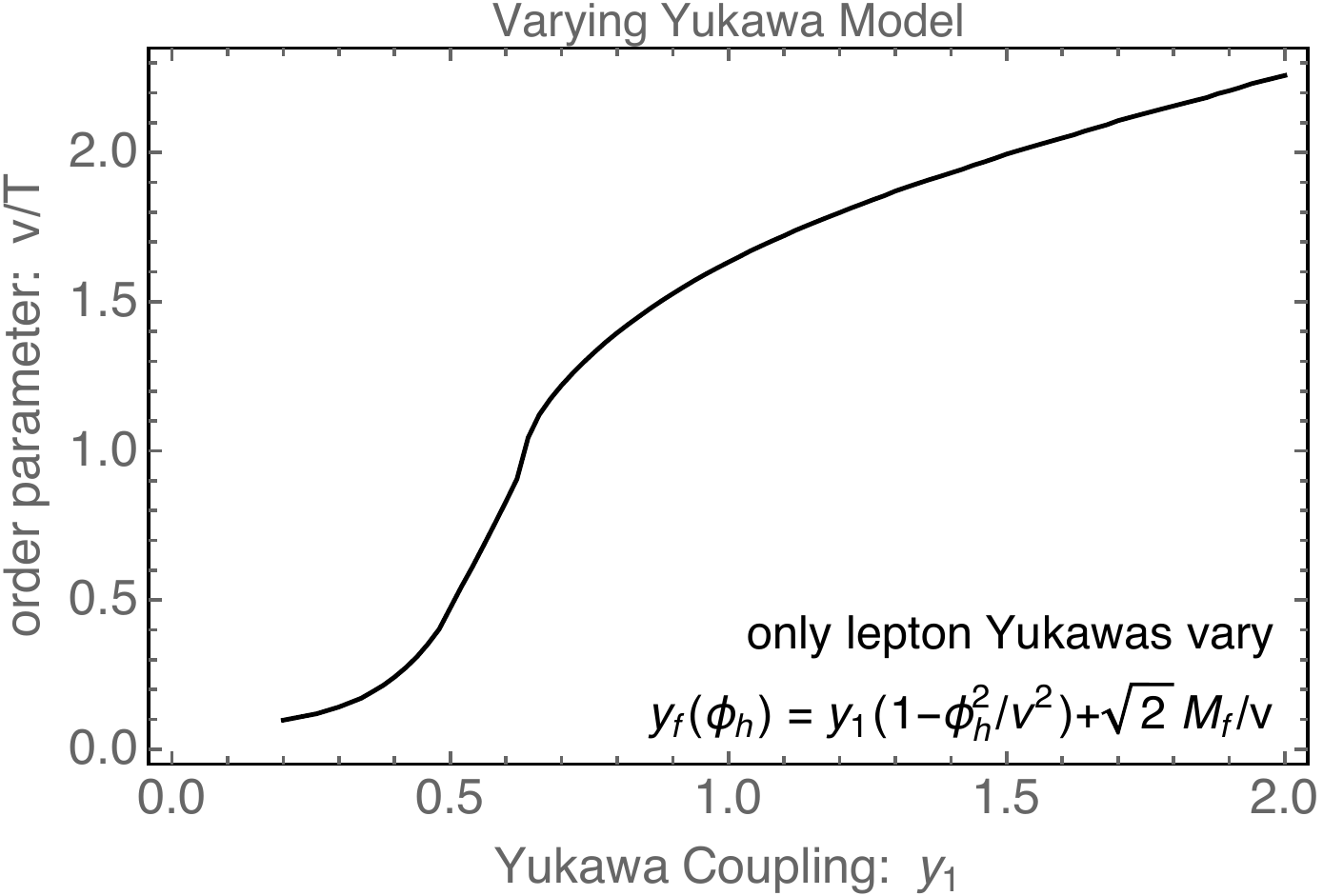} \hfill
\includegraphics[height=6.0cm]{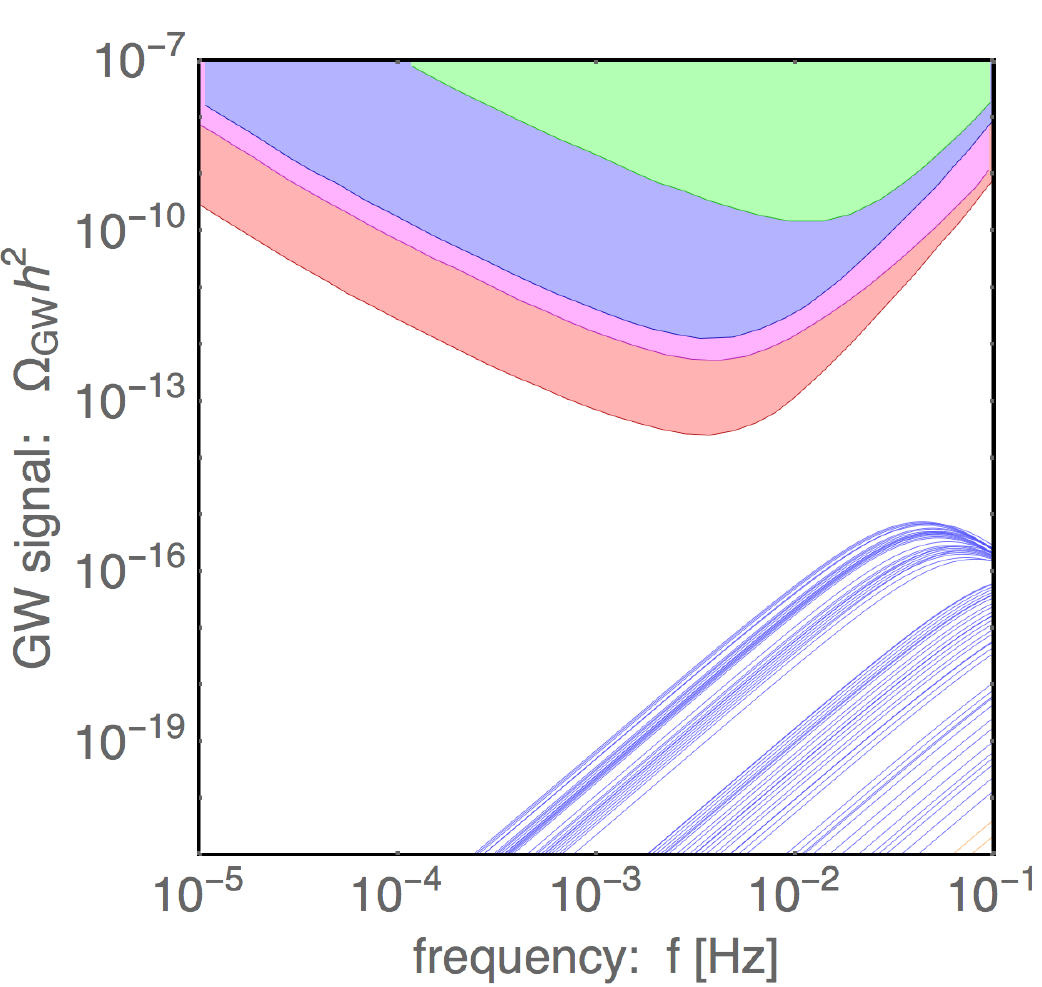} \\ \vspace{0.5cm}
\includegraphics[height=6.0cm]{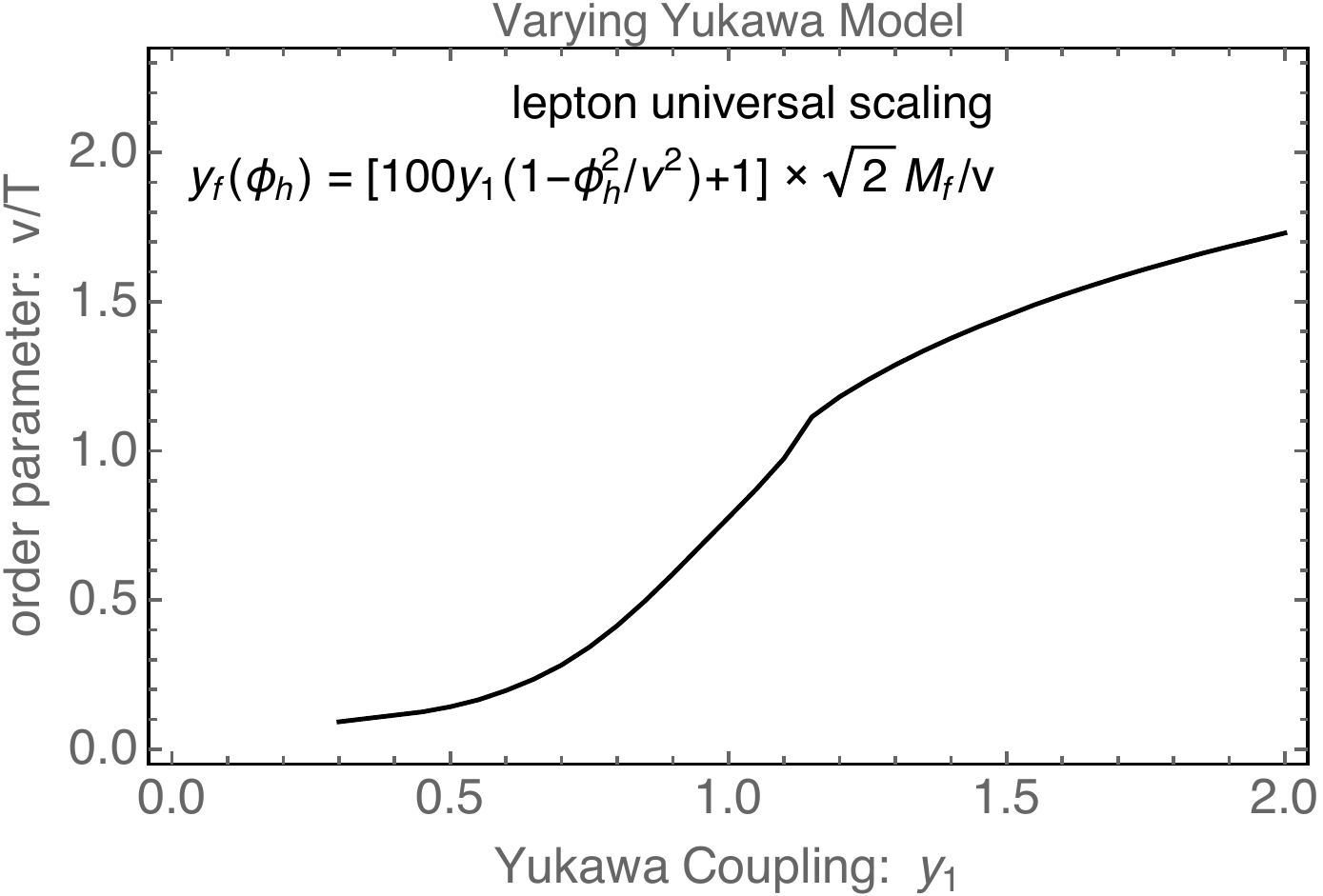} \hfill
\includegraphics[height=6.0cm]{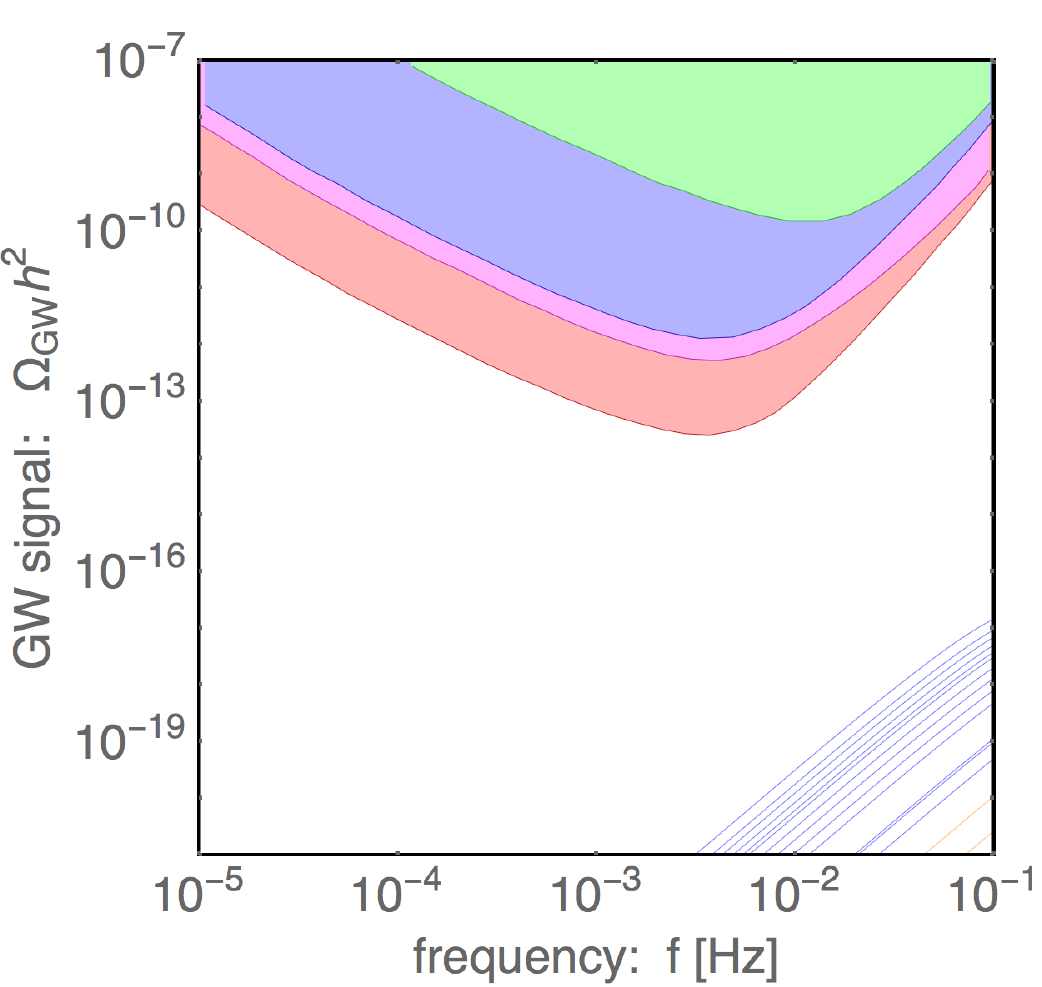} 
\caption{
\label{fig:vary_yukawas}
Results for the varying Yukawas model of \sref{sec:Vary_Yukawa}.  
In the top row, the quark Yukawa couplings are allowed to vary with lepton couplings fixed to their SM values.  In the middle and bottom rows, the lepton couplings vary, and the quark couplings are fixed.  In the first two rows, the Yukawa coupling changes by a flavor-independent shift, and in the bottom row it is a flavor-independent rescaling instead.  
}
\end{center}
\end{figure}

%============
Results of the phase transition analysis appear in \fref{fig:vary_yukawas} where we show the electroweak order parameter $v/T$ in terms of the universal Yukawa parameter $y_1$.  
For cases $(A)$ and $(B)$ above, the electroweak phase transition is strongly first order for $y_1 \gtrsim 0.4$ for varying quark Yukawas and $y_1 \gtrsim 0.7$ for varying lepton Yukawas.  
For the same value of $y_1$ the phase transition is stronger in case $(A)$, because more degrees of freedom have the anomalous field dependence.  
For case $(C)$, the phase transition only becomes strongly first order for $y_1 \gtrsim 1.3$.  
In this model, all the lepton Yukawa couplings are enhanced by the same factor, and consequently the electron and muon remain negligible compared to the tau.  
Effectively, only the one degree of freedom ($\tau$) is playing any role in making the phase transition first order.  
We focus on $y_1 \lesssim 2.0$ to avoid issues associated with loss of perturbatively.  
In this parameter regime, the predicted stochastic gravitational wave background is not within the reach of eLISA's most optimistic design sensitivity.  

%============

%==================================
% CONCLUSION
%==================================
\section{Conclusion}\label{sec:Conclusion}

%============
In this paper we have explored a few minimal extensions of the Standard Model in which new particles below the TeV scale cause the electroweak phase transition to become first order.  
Although the new particle content is motivated from a bottom-up and minimalist perspective, this new physics can easily be embedded in a broader UV theory such as supersymmetry.  
Despite their simplicity, our models exhibit various different mechanisms giving rise to a first order phase transition \cite{Chung:2012vg}.  
For instance, the value of the scalar singlet field can change along with the Higgs field thereby inducing a first order phase transition through tree-level interactions, or the presence of stop-like scalar particles in the electroweak plasma can lead to a first order transition via thermal effects.  
Despite this diversity of particle content, phenomenology, and phase transition dynamics, we find a generic relationship between models with a first order electroweak phase transition and those that are testable at colliders.  
Namely, in the region of parameter space where the electroweak phase transition becomes first order, we typically find such a large deviation in the Higgs coupling with Z-bosons ($hZZ$) that it should be discovered by a future Higgs factory.\footnote{The authors of \rref{Katz:2014bha} have reached similar conclusions.}  
It generically predicts a large, $O(1)$, deviation in the triple Higgs coupling, which can be measured well by next generation hadron collider and high energy lepton collider, such as the $1 \TeV$ version of the ILC. 
The models can also be probed by the Higgs diphoton decay rate, and other methods.  

%============
While future colliders may shed light on the dynamics behind electroweak symmetry breaking, the most direct probe of a cosmological first order phase transition is the associated stochastic background of gravitational waves.  
We have calculated the spectrum of gravitational waves that would have been produced during a first order electroweak phase transition in the early universe.  
We find that only models with especially strong first order phase transitions, typically $v(T)/T \gtrsim 3$, are within reach of a future space-based gravitational wave interferometer experiment like eLISA.  
Nevertheless, models with weaker first order transitions can still be probed by future colliders.  
This provides the exciting opportunity to make complimentary measurements of the electroweak phase transition using a combination of next-generation colliders and interferometers.  
Detecting signals of first order electroweak phase transition in both cosmological observation and collider experiments would be an achievement, which can rival that of the BBN, and open a new page in our understanding of early universe.

%----------------------------------------------------------------
% Acknowledgements
%----------------------------------------------------------------
\quad \\
\noindent
{\bf Acknowledgments:} 
We are grateful to Michael Fedderke for various discussions of the heavy fermion model.  
We thank Jing Shu and Carlos Wagner for interesting discussions.  
We are especially grateful to Jonathan Kozaczuk and Ian Lewis for pointing out a factor of $2$ error in \eref{eq:singlet_dZh}.  
Work at ANL is supported in part by the U.S. Department of Energy under Contract No. DE-AC02-06CH11357. 
P.H. is partially supported by U.S. Department of Energy Grant DE-FG02-04ER41286.
AJL is supported at the University of Chicago by the Kavli Institute for Cosmological Physics through grant NSF PHY-1125897 and an endowment from the Kavli Foundation and its founder Fred Kavli.  
LTW is supported by DOE grant DE-SC0013642.

%----------------------------------------------------------------
%----------------------------------------------------------------
%----------------------------------------------------------------
\appendix

%==================================
% Phase Transition Calculation
%==================================
\section{Phase Transition Calculation}\label{app:PT_calc}

%============
For each of the models, we use the thermal effective potential to calculate the parameters of the phase transition.  
For a pedagogical approach to this calculation, see the lectures in \rref{Quiros:1999jp}.  
For additional details see Appendices~F~and~G of \rref{Chung:2011it}.  
Here we enumerate the various assumptions and approximations.  

%============
We calculate the thermal effective potential in the one-loop approximation.  
The zero-temperature quantum contribution takes the form of the well-known Coleman-Weinberg potential.  
The Coleman-Weinberg potential is dominated by the particles with largest coupling to the Higgs.  
Thus we include the top and anti-top quarks, the weak gauge bosons, the Higgs boson, and model-dependent new physics.  
We neglect the contributions from leptons and lighter quarks, while photons and gluons do not contribute at one-loop order.  
We work in the Landau gauge where the ghosts are massless and do not contribute to the effective potential.\footnote{It is well-known that the effective potential is gauge-dependent \cite{Dolan:1973qd, Nielsen:1975fs} and care must be taken to extract gauge-invariant observables from it \cite{Patel:2011th, Garny:2012cg}.  Since we are not interested in numerically precise results, we instead follow the naive approach outlined in the text.  For a quantitative comparison with the gauge-invariant approach, see {\it e.g.} Refs.~\cite{Wainwright:2011qy, Wainwright:2012zn}.  }
We regulate the one-loop ultraviolet divergences in the dimensional regularization scheme.  
We renormalize at the scale $\mu = M_Z \simeq 91.2 \GeV$ by requiring all possible derivatives of the one-loop correction to vanish at the vacuum.  
That is, we determine the counterterms by imposing $\partial \Delta V_1^0 / \partial \phi_h \bigr|_{\phi_h = v, \phi_s = v_s} = 0$, $\partial^3 \Delta V_1^0 / \partial \phi_h \partial \phi_s^2  \bigr|_{\phi_h = v, \phi_s = v_s} = 0$, and so on.  
(See also Appendix~F of \rref{Chung:2011it}.)  
In this way, the zero-temperature correction to the one-loop thermal effective potential is determined.  
The finite-temperature correction is evaluated from the bosonic and fermionic thermal functions using their representation as a sum over Bessel functions.  
We truncate the sums at $n=5$, which is a good approximation to the full result.  
We neglect the daisy resummation.  
For the models in which the effective potential has a barrier already at $T=0$, such as the singlet models, we expect this to be a good approximation.  

%============
The phase transition is studied by varying the thermal effective potential with respect to temperature.  
For some range of temperatures (possibly including $T=0$) the potential displays a barrier separating a local minimum with $\phi_h = 0$ (symmetric phase) from a local minimum with $\phi_h \neq 0$ (broken or Higgs phase).  
We construct an interpolating trajectory between these local minima (see Appendix~G of \rref{Chung:2011it}) and calculate the $\SO{3}$-symmetric bounce solution and its action $S_3(T)$ along the one-dimensional trajectory.  

%============
The phase transition is said to begin when the bubble nucleation rate $\Gamma_{n}$ exceeds the Hubble expansion rate $H$.  
The condition $\Gamma_{n} > H$ evaluates to 
\begin{align}
	S_{3}(T)/T \lesssim 142
	\com
\end{align}
and we define the bubble nucleation temperature $T_{n}$ such that $S_{3}(T_{n})/T_{n}=142$.  
The phase transition continues until an $O(1)$ fraction of the universe is in the Higgs phase, and we identify this as the percolation temperature $T_{p}$.  
After percolation completes, the energy stored in the bubble walls is transferred back to the plasma as heat, and the temperature increases to $T_{reh} > T_{p}$.  
For simplicity, we will evaluate $T_{n}$ and assume $T_{reh} \approx T_{p} \approx T_{n} \equiv T_{\PT}$.  
For a more sophisticated approach, see \rref{Leitao:2015fmj}.  

%============
The phase transition is characterized by two parameters: $\alpha$ and $\beta/H$.  
In the broken phase at temperature $T_{\PT}$, the vacuum energy $\rho_{{\rm vac},b}$ is calculated as the value of the zero-temperature effective potential at the location of the minimum of the thermal effective potential.  
The vacuum energy in the unbroken phase $\rho_{{\rm vac},u}$ follows from a similar calculation.  
Then the dimensionless parameter 
\begin{align}
	\alpha = \frac{\rho_{{\rm vac},u} - \rho_{{\rm vac},b}}{\rho_{{\rm rad},b}} \Bigr|_{T=T_{\PT}}
\end{align}
quantifies the energy liberated (latent heat).  
Here, $\rho_{{\rm rad},b} = (\pi^2/30) g_{\ast,\PT} T_{\PT}^4$ is the radiation energy density in the broken phase at the temperature of the phase transition.  
If reheating is negligible, as we have assumed, then $0 < \alpha \lesssim 1$, and larger $\alpha$ corresponds to a stronger phase transition with greater supercooling.  
The second parameter $\beta/H$ quantifies the duration of the phase transition.  
During the phase transition, the bubble nucleation rate has an exponential time dependence $\Gamma_n \propto e^{\beta(t-t_0)}$ where $\beta^{-1}$ is the phase transition duration.  
The exponent is calculated as 
\begin{align}
	\frac{\beta}{H_{\PT}} \approx T \frac{d(S_3/T)}{dT} \Bigr|_{T=T_{\PT}}
	\per
\end{align}
This affects the size of bubbles at the time of collision, and therefore it factors into the gravitational wave spectrum.  
We must have $1 \lesssim \beta/H$ otherwise the phase transition would not occur.  

%==================================
% Gravitational Wave Spectrum
%==================================
\section{Gravitational Wave Spectrum}\label{app:GW_calc}

%============
We calculate the spectrum of gravitational waves generated at the electroweak phase transition by following \rref{Caprini:2015zlo}, which summarizes the results of various original sources.  

%============
Gravitational waves at a first order phase transition are produced in three ways: from the collision of bubbles, from the decay of magnetohydrodynamic turbulence, and from the propagation of sound waves.  
The spectrum of gravitational waves produced by bubble collisions is given by the envelope approximation to be 
\begin{align}
	\Omega_{\phi}h^2 & = (1.67 \times 10^{-5}) \left( \frac{\beta}{H_{\PT}} \right)^{-2} \left( \frac{\kappa_{\phi} \alpha}{1+\alpha} \right)^2 \left( \frac{g_{\ast,\PT}}{100} \right)^{-1/3} \left( \frac{0.11 v_w^3}{0.42+v_w^2} \right) \frac{3.8(f/f_{\phi})^{2.8}}{1+2.8(f/f_{\phi})^{3.8}} 
	\com
\end{align}
where $\kappa$ is the fraction of the liberated vacuum energy transferred into motion of the bubble wall and $v_w$ is the speed of the wall.  
The spectrum peaks at a frequency $f_{\phi}$ given by 
\begin{align}
	f_{\phi} & = (1.65 \times 10^{-5} \Hz) \left( \frac{0.62}{1.8-0.1v_{w}+v_{w}^2} \right) \left( \frac{\beta}{H_{\PT}} \right) \left( \frac{T_{\PT}}{100 \GeV} \right) \left( \frac{g_{\ast,\PT}}{100} \right)^{1/6} 
	\per
\end{align}
The decay of magnetohydrodynamic turbulence contributes to the gravitational wave spectrum as 
\begin{align}
	\Omega_{\rm turb}h^2 & = (3.35 \times 10^{-4}) \left( \frac{\beta}{H_{\PT}} \right)^{-1} \left( \frac{\kappa_{\rm turb} \alpha}{1+\alpha} \right)^{3/2} \left( \frac{g_{\ast}}{100} \right)^{-1/3} v_{w} \frac{(f/f_{\rm turb})^{3}}{(1+f/f_{\rm turb})^{11/3}(1+8\pi f/h_{\ast})}
\end{align}
where $\kappa_{\rm turb}$ is the fraction of energy transferred to turbulent motions of the plasma, and 
\begin{align}
	h_{\ast} = \bigl( 1.65 \times 10^{-5} \Hz \bigr) \left( \frac{T_{\PT}}{100 \GeV} \right) \left( \frac{g_{\ast,\PT}}{100} \right)^{1/6}
	\per
\end{align}
This spectrum peaks at a frequency $f_{\rm turb}$ given by 
\begin{align}
	f_{\rm turb} & = (2.7 \times 10^{-5} \Hz) \frac{1}{v_{w}} \left( \frac{\beta}{H_{\PT}} \right) \left( \frac{T_{\PT}}{100 \GeV} \right) \left( \frac{g_{\ast,\PT}}{100} \right)^{1/6} 
	\per
\end{align}
Finally, sound waves contribute to the gravitational wave spectrum as
\begin{align}
	\Omega_{\rm sw}h^2 & = (2.65 \times 10^{-6}) \left( \frac{\beta}{H_{\PT}} \right)^{-1} \left( \frac{\kappa_{v} \alpha}{1+\alpha} \right)^2 \left( \frac{g_{\ast}}{100} \right)^{-1/3} v_{w} \frac{7^{7/2}(f/f_{sw})^{3}}{[4+3(f/f_{sw})^2]^{7/2}}
	\com
\end{align}
where $\kappa_{v}$ is the fraction of energy transferred to bulk motions of the fluid.  
The spectrum peaks at 
\begin{align}
	f_{\rm sw} & = (1.9 \times 10^{-5} \Hz) \frac{1}{v_{w}} \left( \frac{\beta}{H_{\PT}} \right) \left( \frac{T_{\PT}}{100 \GeV} \right) \left( \frac{g_{\ast,\PT}}{100} \right)^{1/6} 
	\per
\end{align}
Note that $\Omega_{\rm turb}h^2$ and $\Omega_{\rm sw}h^2$ are larger than $\Omega_{\phi}h^2$ by a factor of $(\beta/H_{\PT})$.  
This is because gravitational wave production from plasma effects can continue even after bubble collisions have completed.  

%============
The efficiency factors ($\kappa$'s) are model-dependent and vary with the strength of the phase transition.  
If the phase transition is weakly first order, the energy in the scalar field is depleted by interactions with the plasma, and it reaches a terminal velocity (possibly relativistic).  
Then the energy transferred to the plasma ($\kappa_{\rm turb}$ and $\kappa_{v}$) is limited by the vacuum energy that was initially available in the field, which is parametrized by $\alpha$.  
On other hand, for a strongly first order transition the pressure gradient drives the bubble wall to expand and ``run away'' with $v_w \to 1$.  
In this regime, the amount of energy transferred to the plasma saturates, and the surplus energy causes the bubble wall to accelerate.  
Consequently, for large $\alpha$ we have $\kappa_{\phi} \to 1$ and $\kappa_{\rm turb}, \kappa_{\rm sw} \sim 1 / \alpha \to 0$.  

%============
Since only the strongest phase transitions will be detectable by eLISA, we can simplify by assuming that all models lead to runaway phase transitions.  
In this way, we accurately assess the gravitational wave signal for the strongest phase transitions.  
Following \rref{Caprini:2015zlo}, the following expressions parametrize the fraction of energy that is carried by the bubble wall, that is carried by the bulk motion of the plasma, and that is lost as heat:  
\begin{align}
	\kappa_{\phi} = 1 - \frac{\alpha_{\infty}}{\alpha} 
	\ , \quad
	\kappa_{v} = \frac{\alpha_{\infty}}{\alpha} \kappa_{\infty}
	\ , \quad
	\kappa_{\rm therm} = 1 - \kappa_{\phi} - \kappa_{v}
	\per
\end{align}
A fraction $\epsilon$ of the energy carried by bulk motions is transferred to MHD turbulence, $\kappa_{\rm turb} = \epsilon \kappa_{v}$.  
Simulations motivate 
\begin{align}
	\kappa_{\infty} = \frac{\alpha_{\infty}}{0.73 + 0.083 \alpha_{\infty}^{1/2} + \alpha_{\infty}}
\end{align}
and $\epsilon \approx 5\%$.  
We also take $v_{w} = 0.95$.  
These expressions are valid for sufficiently strong phase transitions, $\alpha > \alpha_{\infty}$ where 
\begin{align}
	\alpha_{\infty} \simeq \bigl( 4.9 \times 10^{-3} \bigr) \left( \frac{v(T_{\PT})}{T_{\PT}} \right)^2
	\per
\end{align}
We calculate $T_{\PT}$, $v(T_{\PT})$, $\alpha$, $g_{\ast.\PT}$, and $\beta/H_{\PT}$ from the one-loop thermal effective potential, as described in \aref{app:PT_calc}.  

%----------------------------------------------------------------
% References
%----------------------------------------------------------------
\bibliographystyle{h-physrev5}
\bibliography{EWPT_and_hZZ}

\end{document}